\documentclass[12pt,a4paper]{article}
\usepackage{amssymb}
\usepackage{amsmath}
\usepackage{amsthm}
\usepackage[utf8]{inputenc}
\usepackage{physics}
\usepackage[shortlabels]{enumitem}
\usepackage{mathtools}
\usepackage{bbold}
\usepackage{cite}
\usepackage{adjustbox}
\usepackage{subcaption}
\newcommand{\commentout}[1]{}
\DeclareMathOperator*{\SumInt}{%
\mathchoice%
  {\ooalign{$\displaystyle\sum$\cr\hidewidth$\displaystyle\int$\hidewidth\cr}}
  {\ooalign{\raisebox{.14\height}{\scalebox{.7}{$\textstyle\sum$}}\cr\hidewidth$\textstyle\int$\hidewidth\cr}}
  {\ooalign{\raisebox{.2\height}{\scalebox{.6}{$\scriptstyle\sum$}}\cr$\scriptstyle\int$\cr}}
  {\ooalign{\raisebox{.2\height}{\scalebox{.6}{$\scriptstyle\sum$}}\cr$\scriptstyle\int$\cr}}
}

\usepackage{hyperref}
\hypersetup{
    colorlinks,
    citecolor=blue,
    filecolor=red!75!black,
    linkcolor=blue,
    urlcolor=blue
}

\numberwithin{equation}{section}

\usepackage{tikz-feynman}
\tikzfeynmanset{compat=1.1.0}

\usepackage{appendix}

\usepackage{authblk}
\theoremstyle{definition}

\newcommand{\M}{\mathcal{M}}
\newcommand{\A}{\mathcal{A}}

\newcommand{\E}{\mathcal{E}}
\newcommand{\W}{\mathcal{W}}

\DeclareMathAlphabet{\mathmybb}{U}{bbold}{m}{n}

\begin{document} 

\thispagestyle{empty}

\begin{center}

 \begin{flushright}
 {\small LITP-25-17}\\
 \end{flushright}
\vspace{2cm}

{\Large \bf Positivity with Long--Range Interactions }

\vspace{1.2cm}
{B. Bellazzini${}^{*}$,  J. Berman${}^\dagger$, G. Isabella${}^{\sharp}$, F. Riva$^{\circ}$, M. Romano${}^{*}$, F. Sciotti${}^{\flat}$}

 \vspace*{.5cm} 
{\footnotesize{\textit{\noindent 
 $^{*}$Universit\'e Paris-Saclay, CNRS, CEA, Institut de Physique Th\'eorique, 91191, Gif-sur-Yvette, France.\\ 
 ${}^\dagger$Leinweber Center for Theoretical Physics, Randall Laboratory of Physics, University of Michigan, Ann Arbor, 450 Church St, Ann Arbor, MI 48109-1040, USA\\
 $^{\sharp}${Mani L. Bhaumik Institute for Theoretical Physics, Department of Physics and Astronomy, University of California Los Angeles, Los Angeles, CA 90095, USA}\\
 ${}^\circ$Départment de Physique Théorique, Université de Genève,
24 quai Ernest-Ansermet, 1211 Genève 4, Switzerland
 \\
 ${}^{\flat}$ Departament de F\'isica, Universitat Aut\`onoma de Barcelona, 08193 Bellaterra, Barcelona
}}}
\vspace*{10mm}

\begin{abstract}\noindent\normalsize

We introduce infrared finite, analytic, crossing symmetric, Regge behaved, and Lorentz invariant amplitudes $\M_{\E}$, labeled by the experimental energy resolution $\E$ for detecting soft photons and gravitons. For $\E$ exponentially smaller than any hard scale, they also satisfy unitarity and their associated cross sections reproduce the inclusive, infrared–finite cross sections of ordinary amplitudes. These properties make $\M_{\E}$ suitable for deriving infrared–safe positivity bounds on effective field theories in the presence of long–range forces even in $D=4$. As an illustration, we present explicit bounds in the low–energy theory of pions coupled to electromagnetism and gravity.

\end{abstract}

\end{center}

\newpage
\clearpage
\pagenumbering{Roman}

\renewcommand{\baselinestretch}{0.9}\normalsize
{ \hypersetup{hidelinks} 
 \tableofcontents }
\renewcommand{\baselinestretch}{1}\normalsize

\newpage
\clearpage
\pagenumbering{arabic}

\section{Introduction}
\label{Sec:Intro}

Scattering amplitudes lie at the core of both theoretical and experimental high-energy physics. 
They  connect directly to physical observables accessible in experiments and at the same time encode the principles of quantum mechanics and relativity in a remarkably compact form. %Amplitudes are often regarded as fundamental objects in their own right, elevated to the status of primitive elements of the theory.  
For this reason, amplitudes play a central role in the S-matrix bootstrap program, through which causality and unitarity can define the space of UV-completable effective field theories (EFTs) through positivity constraints, see e.g.~\cite{Adams:2006sv,Camanho:2014apa,Pham:1985cr,Pennington:1994kc,Ananthanarayan:1994hf,Distler:2006if,Vecchi:2007na,Manohar:2008tc,Bellazzini:2014waa,Bellazzini:2015cra,Bellazzini:2016xrt,Cheung:2016yqr,Bonifacio:2016wcb,Hinterbichler:2017qcl,deRham:2017avq,deRham:2017zjm,Bellazzini:2017fep,Bellazzini:2018paj,Bellazzini:2019bzh,Bellazzini:2019xts,Alberte:2020bdz,Englert:2019zmt,Melville:2019tdc,Bellazzini:2020cot,Arkani-Hamed:2020blm,Caron-Huot:2020cmc,Tolley:2020gtv,Bellazzini:2021oaj,Bern:2021ppb,Chiang:2021ziz,Caron-Huot:2021rmr,Chala:2021wpj,Davighi:2021osh,Arkani-Hamed:2021ajd,Bellazzini:2021shn,Serra:2022pzl,Caron-Huot:2022jli,Bellazzini:2022wzv,Caron-Huot:2022ugt,Guerrieri:2022sod,Henriksson:2022oeu,Haring:2022sdp,Creminelli:2022onn,Bellazzini:2023nqj,Karateev:2023mrb,Haring:2023zwu,McCullough:2023szd,Albert:2023jtd,Ma:2023vgc,Berman:2023jys,EliasMiro:2023fqi,Albert:2023seb,Creminelli:2023kze,Berman:2024kdh,Albert:2024yap,Berman:2024eid,Berman:2024owc,Creminelli:2024lhd,Dong:2024omo,Bertucci:2024qzt,Cremonini:2024lxn,Remmen:2024hry,Haring:2024wyz,Chakraborty:2024ciu,Tokareva:2025rta,Cheung:2025krg,deRham:2025vaq,Peng:2025klv,Chang:2025cxc,Hui:2025aja,Correia:2025uvc,Ye:2025zhs,Berman:2025owb,Bellazzini:2025shd,Bonnefoy:2025uzf,Dong:2025dpy,Creminelli:2025rxj,Huang:2025icl}.

This perspective is challenged in four spacetime dimensions by the existence of long-range forces mediated by photons and gravitons, whose infrared (IR) divergences trivialize scattering amplitudes, driving them to zero.\footnote{An equivalent  viewpoint on IR divergences has emerged in the context of memory effect \cite{Satishchandran:2019pyc,Prabhu:2022zcr} and asymptotic symmetries---large gauge transformations in QED and BMS symmetries in gravity---where amplitudes vanish because they violate selection rules associated with asymptotic charges~\cite{Gabai:2016kuf,Kapec:2017tkm,Strominger:2017zoo}.} 
IR divergences are tamed in inclusive observables, as guaranteed by the Bloch–Nordsieck and KLN theorems~\cite{Bloch:1937pw,Kinoshita:1962ur,Lee:1964is}, as well as in certain IR-finite quantities such as energy–energy correlators~\cite{Moult:2025nhu}. Yet these observables have not been successfully incorporated into the bootstrap program, leading to an apparent breakdown of amplitude constraints as soon as long-range mediators are present, no matter how weakly they couple. 
Although this discontinuity seems paradoxical, it is natural to wonder whether such extreme UV/IR mixing, perhaps reminiscent of the swampland program~\cite{Vafa:2005ui,ArkaniHamed:2006dz,Palti:2019pca}, 
might be unavoidable.

In this paper, we address this specific question, showing that one can construct bounds that smoothly interpolate between finite and vanishing couplings in electromagnetism and gravity.
The key observation is that
long-range interactions  require initial and final scattering states to be specified, at least partially, in terms of properties of an experimental apparatus. Unlike in the case of short-range interactions or  IR-safe observables, the apparatus cannot be fully decoupled from the scattering process under study.   
This is reflected, for example, in the infinitely many viable choices of asymptotic Hamiltonians and dressings in the Faddeev--Kulish approach to IR-finite amplitudes~\cite{Kulish:1970ut,Kapec:2017tkm,Hannesdottir:2019opa}.

We focus on experimental apparatuses of \emph{finite} proper size $\E^{-1}$ which, in line with the traditional definition of amplitudes, we take exponentially larger than any intrinsic distance-scale $1/M$ of the system under study: $\log(M/\E)\gg1$. 
The ability of the apparatus to detect low-energy (soft) particles is limited by its finite inverse size.  
In these conditions, the most important effects 
are captured by  the scaling limit,
\begin{equation}
\label{eq:generalscaling}
\begin{split}
&\alpha\to 0\,,\,\,\,\E\to 0\,,\quad   \alpha_{\E}\equiv\alpha\log M/\E =\mathrm{fixed} \qquad \mathrm{(electromagnetism)} \\
&G\to 0\,,\,\,\,\E\to 0\,,\quad   {G}_\E\equiv G M^2\log M/\E =\mathrm{fixed} \qquad \mathrm{(gravity)}
\end{split}
\end{equation}
with $G=1/8\pi m^2_{\mathrm{Pl}}$ the Newton constant,  $m_{\mathrm{Pl}}$ the reduced Planck mass, and $\alpha=e^2/4\pi$ the fine structure constant. 
Crucially, the theory effectively depends on \emph{two couplings}, $\alpha$ and $\alpha_{\E}$ ($G M^2$ and $G_{\E}$) ---in addition to any other short--range interactions--- and these couplings play qualitatively and quantitatively  separate roles.
The  scaling limit   isolates the {\it universal} long--distance effects common to all sufficiently large detectors, in terms of  a single parameter,~$\alpha_{\E}$~($G_{\E}$). 
%
%The finite experimental resolution $\E$
Finite corrections from the couplings $\alpha$ and $G$, as well as finer
dependence on $\E$ and other details of the experimental apparatus, provide
subleading contributions to observables, entering only at next--to--leading
order in the scaling~\eqref{eq:generalscaling}.~\footnote{Perhaps the famous quote by V.I. Arnold  ``Mathematics is the part of physics where experiments are cheap'' \cite{Arnold:98} helps expressing our perspective on the physical interpretation of the mathematical limit \eqref{eq:generalscaling}.   }

We then define \emph{stripped amplitudes} 
 $\M_{\E}$ as
a formal amplitude in the finite-size experiment, by factoring out the analytic extension of Weinberg IR--divergent exponentials \cite{Weinberg:1965nx}. We prove that the stripped amplitudes are IR-finite, analytic, crossing symmetric, Lorentz invariant, and Regge--behaved. We moreover show that they are unitary in the scaling limit~\eqref{eq:generalscaling}, and that their
exclusive cross sections reproduce the IR--finite inclusive ones controlled by the
energy–resolution scale~$\E$.
The universality of this leading behavior also ensures that in the same scaling limit they match Faddeev--Kulish amplitudes. 

These features allow dispersive methods to be extended to  theories with long-range interactions and  derive positivity bounds on EFT Wilson coefficients from dispersive sum rules for $2\to2$ stripped amplitudes. 
As explicit examples, we establish IR--finite positivity bounds in the EFT of pions coupled to electromagnetism or gravity. 
Notably, these are also free  of the issues that arise due to the photon/graviton $1/t$-pole in four dimensions~\cite{Caron-Huot:2021rmr,Henriksson:2022oeu}.
Moreover, they resolve the puzzle of non-decoupling gravitational negativity~\cite{Chang:2025cxc}, showing that only a finite, decoupling contribution $\sim GM^2\log M/\E$ remains, while robust inequalities hold in $D=4$.

As a simple illustration, consider the EFT of a single pion with  $2\to2$ scattering amplitude  $\M=\hat{c}_{2,0}(s^2+t^2+u^2)/2+\ldots$.   In the absence of long--range forces, positivity implies the familiar bound $\hat{c}_{2,0}\ge0$.  In the presence of gravity, the bound takes the following form
\begin{equation}
\A_0(\hat{c}_{n,k},G_\E)+O(G)\ge 0\,, 
\end{equation}
with $\A_0$ a known function of Wilson coefficients and of $G_\E$.  
In other words, we obtain \emph{finite} and \emph{calculable} corrections in $D=4$ to the usual positivity constraints, unlike previous analyses. 
For instance, a representative (and not optimized) bound obtained in our analysis reads $ \hat{c}_{2,0}M^4+65.7\!\left(8\pi G_{\E}+O(G_{\E}^2)\right)+O(G)>0$, 
with $G_{\E}$ encoding the universal dominant long--range effect of gravity, which is systematically calculable (see~\eqref{eq:boundsgr1} and~\eqref{eq:boundsgr2} for the explicit $O(G_{\E}^2)$ terms).  
The smallness of  $G/G_{\E}\ll1$ ensures parametric control over the $O(G)$ corrections.

The paper is organized as follows. In Section~\ref{Sec:introducingStripped} we introduce the stripped amplitudes $\M_\E$, then study their analytic continuation and basic properties in Section~\ref{section:Analytic StrippedAmplitudes}. We discuss unitarity in Section~\ref{section:unitarity}, and derive dispersion relations together with positivity bounds in Section~\ref{sect:dispersion_relations}. We present explicit applications to the theory of pions coupled to electromagnetism and gravity in Sections~\ref{Sec:BoundsEM} and~\ref{sec:positivitygravity}, respectively. Finally, we give a comparison with previous literature and our conclusions in Section~\ref{sect:conclusions} and collect technical derivations and ancillary material in the Appendices.

\section{Stripped and Hard Amplitudes}
\label{Sec:introducingStripped}

Consider a relativistic quantum field theory in flat spacetime \emph{without} long-range forces, but  possessing a conserved energy--momentum tensor and possibly  a conserved $U(1)$ current.
We perturb the system in two ways: {\it (i)} first by turning on long-range interactions mediated by dynamical photons, gravitons, or both; and then {\it (ii)} by introducing an IR regulator $\epsilon$, which, 
for definiteness, we adopt to be a dimensional regulator,\footnote{
Alternatives, such as a finite photon mass $m_{\gamma}$ (with photons coupled to conserved currents and removing  terms such as $m^4_\gamma A_\mu^4$, to avoid  nontrivial longitudinal scattering in the $m_\gamma\!\to\!0$ limit) or a hard cutoff $\lambda$ on three-momenta $|\boldsymbol{k}|>\lambda$, are equally admissible, since all regulators are removed in the end.
These schemes are recovered from dimensional regularization by replacing in the Weinberg exponentials $1/\epsilon\!\to\!\log(\mu^2/m_\gamma^2)$ or $\log(\mu^2/\lambda^2)$, where $\mu$ is the usual renormalization scale.
} 
\begin{equation}
\epsilon=\frac{D-4}{2}>0
\,.
\end{equation}
The regulator ensures that the perturbed amplitudes remain nontrivial and continuously connected to those of the original short-range theory. Concretely, taking $\alpha\!\to\!0$ ($G\!\to\!0$) at fixed $\epsilon$ yields finite amplitudes, while sending $\epsilon\!\to\!0$ afterward recovers the unperturbed limit. In the opposite order, instead, scattering amplitudes vanish.

We next introduce a Lorentz invariant scale $\E>0$, which can correspond to the inverse proper size of an experiment, or its (proper) energy resolution, or even just a reference mass of  a particle. To isolate the most important IR effects,  we momentarily  pick the rest frame of the experiment and label photons and gravitons ``soft''  if they have energy below  $\E$ or ``hard'' otherwise.\footnote{This intermediate step is effectively defining the Lorentz breaking hard amplitude $\M^{\mathrm{hard}}_\E$ that is introduced in \eqref{eq:hardsubleadingseparation}, from which we then abstract to the Lorentz covariant definition of stripped amplitudes in \eqref{eq:Mepslambda} or  \eqref{eq:LambdamplitudesDefQED}.}
%We next introduce a scale $\E>0$ and call photons and gravitons ``soft'' if they have energy below $\E$ or ``hard'' otherwise. 
As the IR regulator $\epsilon$ is removed, the probability for
real soft emission increases, and
the survival probability 
of  exclusive amplitudes among states with only hard particles, $|\boldsymbol{k}|>\E$, 
decays exponentially. In practice, this suppression arises from loops of virtual soft particles with $|\boldsymbol{k}|<\E$, whose singular contributions can be computed, since they exponentiate.
In QED, the exclusive (regulated) amplitude $\M^{\epsilon}$ decays exponentially with the overall factor obtained by adapting the original result by Weinberg~\cite{Weinberg:1965nx},
\begin{equation}
\label{eq:WeinbergExpQED}
 \W^{\mathrm{QED}}%_{\epsilon,\E}
 \equiv \mathrm{exp}\left\{\frac{\alpha }{8\pi}\frac{\left(\E/\mu\right)^{2\epsilon}}{\epsilon}\left(\sum_{i,j} \eta_i \eta_j q_i q_j \frac{1}{\beta_{ij}}\log\frac{1+\beta_{ij}}{1-\beta_{ij}}-i\sum_{\substack{i\neq j \\ \eta_i \eta_j=1}} q_i q_j \frac{2\pi}{\beta_{ij}}\right)\right\} \,,
\end{equation}
where $\eta_i=\pm1$ for {\it in} ($-$) and {\it out} ($+$) going states, $q_i$ are the in/out electric charges,\footnote{While this expression is for in/out notation, in Section~\ref{section:Analytic StrippedAmplitudes} we will compute its analytic continuation, which will enable us to work with all-incoming notation. } and 
$\beta_{ij}\equiv[1-(m_i m_j/k_i\!\cdot\! k_j)^2]^{1/2}$ the relative velocity between particles $i$ and $j$; $\mu$ is the dimensional-regularization scale, which eventually drops out of all physical expressions. Importantly, in the definition of the overall factor $\W^{\mathrm{QED}}$ we singled out the universal Lorentz invariant contribution, despite the intermediate steps involved a frame--dependent labeling for photons.
%and the $O(\epsilon)$ term  is $(\E/\mu)$-independent.
As advertised, $\lim_{\epsilon \to 0^+}\W^{\mathrm{QED}}%_{\epsilon,\E} 
= 0$, while if we first take $\alpha\to 0$ the amplitude is finite as  $\epsilon\to0$.

Gravitational effects lead to a similar exponential suppression~\cite{Weinberg:1965nx},
\begin{equation}
\label{eq:WeinbergExpGR}
\!\!\W^{\mathrm{GR}}%_{\epsilon,\E} 
\! \equiv \!\mathrm{exp}\left\{\! -\frac{G}{8\pi} \frac{\left(\E/\mu\right)^{2\epsilon}}{ \epsilon} \!\!\left(\sum_{i,j} \eta_i \eta_j  m_i m_j \frac{1+\beta^2_{ij}}{\beta_{ij}\sqrt{1-\beta^2_{ij}}}\log\frac{1+\beta_{ij}}{1-\beta_{ij}}-i\!\!\sum_{\substack{i\neq j \\ \eta_i \eta_j=1}}  m_i m_j \frac{2\pi(1+\beta^2_{ij})}{\beta_{ij}\sqrt{1-\beta^2_{ij}}}\!\!\right)\!\!\right\},
 %\xrightarrow[\epsilon \to 0^+]{} 0
\end{equation}
which also vanishes as $\epsilon\!\to\!0^+$.
 
Motivated by the universal factorization of IR--divergent soft contributions, we define the following ratio obtained by stripping off  $\W$ (i.e. $\W^{\mathrm{QED}}$, $\W^{\mathrm{GR}}$ or both) from the full amplitude:
%Motivated by the universal factorization of IR--divergent soft contributions, we isolate the contribution from processes with hard kinematics, $|s_{ij}|>\E^2$, by first defining the ratio obtained by stripping off  $\W$ (i.e. $\W^{\mathrm{QED}}$, $\W^{\mathrm{GR}}$ or both) from the full amplitude:
\begin{equation}\label{eq:Mepslambda}
\M_{\E}\equiv\lim_{\epsilon\to0^+}\M_{\E}^{\epsilon} \,\quad \textrm{with} \quad \M_{\E}^{\epsilon}\equiv\frac{\M^{\epsilon}}{\W%_{\epsilon,\E} 
}\, .
\end{equation}
Since, by construction, $\M_{\E}^{\epsilon}$ is IR finite, the \emph{stripped amplitude} $\M_{\E}$ depends only on the physical scale $\E$ and  kinematics.  

% In perturbation theory,
% $\M_{\E}$ receives contributions  from loops of virtual photons/gravitons that are either  hard $|\boldsymbol{k}|> \E$, or  soft   with $|\boldsymbol{k}|< \E$ but \emph{finite}  (since the singular part has been canceled by $\W$). We write these as

In order to understand the meaning of $\M_{\E}$, it is convenient to pick again a frame and reintroduce the splitting between hard and soft modes. Following the definition of $\W$, $\M_{\E}$ receives contributions from loops of virtual photons/gravitons that are either hard $|\boldsymbol{k}|> \E$, or  soft   with $|\boldsymbol{k}|< \E$ but \emph{finite}  (since the singular part has been canceled by $\W$). We write these as\footnote{\label{ftntLorentz}$\M_\E^{\text{hard}}$ and $\delta_{\text{sub.~soft}}$ are independently frame-dependent, though their sum $\M_{\E}$ is Lorentz-invariant.  The hard--amplitude is approximately invariant w.r.t boosts relating two frames as long as the boost--rapidity is much smaller than $M^2/\E^2$. Moreover, its leading $\alpha_\E$ ($G_\E$) dependence is exactly Lorentz invariant, the only source of breaking being  subleading terms as $\E\to 0$, see Eq. \eqref{eq:m0ismhard}.  } 
\begin{equation}\label{eq:hardsubleadingseparation}
\M_{\E}=\M^{\mathrm{hard}}_{\E}+\delta_{\text{sub. soft}}\, .
\end{equation}
The {\it hard amplitude}  $\M^{\mathrm{hard}}_{\E}$ in \eqref{eq:hardsubleadingseparation}  can be independently defined by a path integral in which the original Lagrangian $\mathcal{L}(A^{\mu},h^{\mu\nu},\phi_i)$ is replaced by $\mathcal{L}(A^{\mu}_{\E},h^{\mu\nu}_{\E},\phi_i)$, where the gauge and gravitational fields $A^{\mu}_{\E}$ and $h^{\mu\nu}_{\E}$ contain no Fourier modes with $|\boldsymbol{k}|<\E$. In particular, all photon and graviton propagators used to perturbatively build $\M^{\mathrm{hard}}_{\E}$ come with a step function $\theta(|\boldsymbol{k}|-\E)$, and the associated interaction doesn't include vertices with soft photons/gravitons. 
On the other hand, $\delta_{\mathrm{sub.~soft}}$ contains  the  contributions   from soft photon or graviton loops which are not captured by the overall divergent exponential $\W$ because they are IR-finite and subleading in the $\E\to 0$ limit.  
Since the IR regulated amplitude $\M^\epsilon$ is independent of $\E$, the stripped amplitude $\M_{\E}$ depends only on $\log\E$ via $\mathcal{W}$, whereas $\M_{\E}^{\mathrm{hard}}$ and  $\delta_{\mathrm{sub.~soft}}$ may separately depend also on powers of $\E/E\ll1$, which cancel out in their sum.\footnote{Ref.~\cite{Weinberg:1965nx} implies $\M^{\epsilon}=\M^{\mathrm{hard}}_{\E} \W+\text{subleading soft}$. However, all evidence from exactly solvable models \cite{Bloch:1937pw,Blanchard:1969} as well as from conservation of asymptotic charges  \cite{Strominger:2017zoo} and path integral resummations \cite{White:2011yy}, indicates that the exponential $\W$ multiplies the full amplitude as a global suppression factor, rather than acting only as a prefactor to the hard part. We adopt this physically safe assumption throughout this work.}
\\

In the construction so far,  $\E$ appears as an arbitrary parameter in $\M_\E$. 
Physically, $\E$ can be identified with the experimental energy resolution $\Delta E$ for detecting real photons or gravitons, \emph{encoded at amplitude-level}.
Indeed, \emph{exclusive} differential cross-sections $d\sigma_{\alpha_\E}/d\Omega_{hard}$ computed from $\M_\E$  are not only IR finite but also reproduce physical \emph{inclusive} cross-sections with resolution $\Delta E=\E$.  For instance in QED, scattering into a hard region of phase space,  $d\Omega_{hard}$, in the scaling limit \eqref{eq:generalscaling} gives, 
\begin{equation}
\label{eq:inclusiveXsection}
\frac{d\sigma_{\alpha_\E}/d\Omega_{hard}}{%d\sigma^\epsilon/d\Omega +
\sum_{n=0}^{\infty} d\sigma^\epsilon_{+n,
\gamma}(E_\gamma<\Delta E)/d\Omega_{hard}}\longrightarrow \left(\frac{\Delta E}{\E}\right)^{-\frac{\alpha }{8\pi^2}\left(\sum_{i,j} \eta_i \eta_j q_i q_j \frac{1}{\beta_{ij}}\log\frac{1+\beta_{ij}}{1-\beta_{ij}}\right) }=_{\big|_{\E=\Delta E}} \!\!\!\!\!\!\! 1\,,
\end{equation}
and analogously in gravity. 
That $\E$ is an arbitrary scale simply reflects the freedom to consider detectors with different sensitivities $\Delta E=\E$. In this limit, and anticipating the amplitude--like properties derived in the next sections, the stripped amplitudes naturally behave as {\it ``detector amplitudes''} associated with a given experimental resolution.

The stripped amplitude  can also be understood as an \textit{equivalence class} of amplitudes, valid in the limit \eqref{eq:generalscaling}.  Amplitudes $\M^\epsilon_{+n\,
\gamma}$ between states with the same hard particles and any number $n$ of real soft photons (or gravitons) are equivalent in the sense that,
\begin{equation}
\lim_{\substack{\alpha\to 0 \\ \E \to 0 \\ \alpha\log(M/\E)\to \alpha_{\E}}}\lim_{\epsilon\to0^+}\frac{\M^\epsilon_{+n\,
\gamma}}{\W^{\mathrm{QED}}\times (n\text{ real soft photon factors})}=\lim_{\substack{\alpha\to 0 \\ \E \to 0 \\ \alpha\log(M/\E)\to \alpha_{\E}}}\M_{\E} \,,\quad \quad\forall n\,.
\end{equation}
This interpretation morally connects the stripped amplitudes to other constructions of IR finite amplitudes, such as those of Faddeev-Kulish, see Section \ref{subsect:Faddeev-Kulish}.

\section{Analytic Stripped Amplitudes}
\label{section:Analytic StrippedAmplitudes}

In this section we extend the stripped amplitudes $\M_{\E}$ from the physical region into analytic functions and prove their basic properties---analyticity, crossing symmetry, Lorentz invariance and  Regge scaling. Along with unitarity, which we discuss in the next section, these properties  qualify them as genuine (representatives of an equivalence class of) amplitudes. 
\subsection{Analytic soft exponentials}\label{sec:analyticsoft}

The exponentials \eqref{eq:WeinbergExpQED} and \eqref{eq:WeinbergExpGR} can be analytically continued into analytic functions of the Mandelstam invariants $s_{ij}$.  
To do so, we first 
factor the IR exponentials into a product
\begin{equation}
\prod_{i,j}\W^{\mathrm{QED}}_{ij}(s_{ij})=\W^{\mathrm{QED}}\,.
\end{equation}
The kinematic dependence of the exponents is contained entirely in the relative velocities $\beta_{ij}$, which, in the physical region, can be expressed as
\begin{equation}
\beta_{ij}^2=\frac{\left[s_{ij}-(m_i-m_j)^2\right]\left[s_{ij}-(m_i+m_j)^2\right]}{(s_{ij}-m_i^2-m_j^2)^2}\,, 
\quad 
m_i m_j \frac{1+\beta_{ij}^2}{\sqrt{1-\beta_{ij}^2}}=\frac{(s_{ij}-m_i^2-m_j^2)^2-2m_i^2 m_j^2}{s_{ij}-m_i^2-m_j^2}\,.
\end{equation}
%Therefore, as $s_{ij} \to \infty$, the exponent goes to a constant, meaning $\W^{\mathrm{QED}}_{ij}/s_{ij}\to0 $ in this limit. 
Therefore, as $s_{ij} \to \infty$, the $\W^{\mathrm{QED}}_{ij}/s_{ij}\to0 $. We can then use once-subtracted dispersion relations to analytically continue the Weinberg exponentials,\footnote{An alternative route is going through a worldline representation, see \cite{Lippstreu:2025jit}. }
\begin{equation}
\W^{\mathrm{QED}}_{ij}(s_{ij})=\W^{\mathrm{QED}}_{ij}(s_0)+\int_{(m_i+m_j)^2}^{\infty}\frac{d\bar{s}}{2\pi i}\frac{s_{ij}-s_0}{(\bar{s}-s_0)(\bar{s}-s_{ij})}\mathrm{Disc}\,\W^{\mathrm{QED}}_{ij}(\bar{s})\,,
\end{equation}
matching the discontinuity from the imaginary part of~\eqref{eq:WeinbergExpQED}, restricted to $s\geq (m_i+m_j)^2$, corresponding to the physical region only (defined in the usual way as the real axis approached from above). We find for complex $s$, 
in all--incoming convention,\footnote{Charges $q_i$ are in all--incoming convention, differently from the crossing--unfriendly version \eqref{eq:WeinbergExpQED}. Analyticity and crossing are discussed in Section~\ref{subsec:properties}.  The function $\arctan(z)$ is defined as $\frac{i}{2}\left[\log(1-iz)-\log(1+i z)\right]$ and has branch cuts along the imaginary axis in the complex $z$--plane. Both $\log z$ and $\sqrt{z}$ are understood as being on their principal branches. }
\begin{equation}
\label{eq:AnalyticSoftQED}
 \W^{\mathrm{QED}}_{ij}(s_{ij}) \equiv \left\{
 \begin{array}{lr}
\mathrm{exp}\left[-\frac{\alpha q_i q_j}{2\pi }\frac{\left(\E/\mu\right)^{2\epsilon}}{\epsilon} \frac{s_{ij}-m_i^2-m_j^2}{\sqrt{(s_{ij}-(m_i-m_j)^2)(-s_{ij}+(m_i+m_j)^2)}} \arctan\frac{\sqrt{s_{ij}-(m_i-m_j)^2}}{\sqrt{-s_{ij}+(m_i+m_j)^2}}\right] & i\neq j \\
 & \\
\mathrm{exp}\left[\frac{\alpha q_i^2}{4\pi}\frac{\left(\E/\mu\right)^{2\epsilon}}{\epsilon} \right] &  i=j
 \end{array}
 \right.
\end{equation} 
which are, by construction, analytic functions with no branch point at the pseudothreshold $(m_i-m_j)^2$, nor at any unphysical location.  For real kinematics, their product matches \eqref{eq:WeinbergExpQED}.
For illustration,  with $m_i^2=m_j^2=m^2\ll |s_{ij}|$, these factors simplify to,
\begin{equation}\label{epsaps}
\W^{\mathrm{QED}}_{ij}(s_{ij})\to \mathrm{exp}\left[\frac{\alpha q_i q_j%\eta_i\eta_j
}{4\pi}\frac{\left(\E/\mu\right)^{2\epsilon}}{\epsilon}\log\left(-\frac{s_{ij}}{m^2}\right)\right]\,,\quad\quad (m^2\ll |s_{ij}|\,, i\neq j)\,.
\end{equation}

We can find a similar analytic continuation in gravity, but with a twice-subtracted dispersion relation,
\begin{equation}
\label{eq:AnalyticSoftGR}
 \W^{\mathrm{GR}}_{ij}(s_{ij}) \equiv \left\{
 \begin{array}{lr}
 \mathrm{exp}\left[\frac{G}{2\pi}\frac{\left(\E/\mu\right)^{2\epsilon}}{\epsilon} \frac{(s_{ij}-m_i^2-m_j^2)^2-2m_i^2 m_j^2}{\sqrt{(s_{ij}-(m_i-m_j)^2)(-s+(m_i+m_j)^2)}} \arctan\frac{\sqrt{s_{ij}-(m_i-m_j)^2}}{\sqrt{-s_{ij}+(m_i+m_j)^2}}\right] & i\neq j \\
 & \\
\mathrm{exp}\left[-\frac{G m_i^2}{4\pi}\frac{\left(\E/\mu\right)^{2\epsilon}}{\epsilon} \right] &  i=j\,.
 \end{array}
 \right.
\end{equation}  
For  vanishing masses $m_i=m_j=0$, the expression simplifies to,
\begin{equation}
\label{eq:WeinbergEqualvanishingMasses}
 \W^{\mathrm{GR}}_{ij}(s_{ij})\Big|_{m_i=m_j=0} = 
 \begin{array}{lr}
 \mathrm{exp}\left[-\frac{Gs_{ij}}{4\pi}\frac{\left(\E/\mu\right)^{2\epsilon}}{\epsilon} \log\left(\frac{-s_{ij}}{\mu^2}\right) \right]
 \end{array}
\end{equation}  
for $i\neq j$ and 1 for $i=j$. Here the $\mu^2$ within the logarithm cancels in the product $\prod_{ij} \W^{\mathrm{GR}}_{ij}(s_{ij})$ due to momentum conservation, ensuring that no collinear singularity arises in gravity --  in contrast with   the QED expression  \eqref{epsaps}.

\subsection{Properties of Stripped Amplitudes}
\label{subsec:properties}

Having extended analytically the Weinberg exponentials, we can now focus on the analytic version of stripped amplitudes following Eq.~\eqref{eq:Mepslambda}:
\begin{equation}
\label{eq:LambdamplitudesDefQED}
\M_{\E} \equiv \lim_{\epsilon\to 0^+} \frac{\M^{\epsilon}}{\prod_{ij}\W^{\mathrm{QED(GR)}}_{ij}(s_{ij}) }\,, \quad\quad
%\label{eq:LambdamplitudesDefGR}
\end{equation}
for electromagnetism or  gravity respectively --  when both interactions are present, both soft exponentials are present. These expressions reduce for physical kinematics to \eqref{eq:Mepslambda}. 

The (analytic version of the) stripped amplitude $\M_{\E}$ satisfies the following properties:

\begin{enumerate}[(a)]

\item {\bf IR finiteness.}  
By construction, factoring out $\prod_{ij} \W_{ij}$ from $\M^\epsilon$ removes all IR divergences. 
Moreover, as long as we work with massive charged particles,
collinear singularities are also absent.
In particular, IR-finite physical observables can be computed directly in terms of $\M_{\E}$.

\item {\bf Lorentz (little-group) covariance.}  
This follows immediately, since $\W_{ij}$ is a Lorentz scalar. In particular, the external momenta enter in $\W_{ij}$ in the Lorentz invariant combination $s_{ij}$.

\item {\bf Analyticity and crossing symmetry.}  
Analyticity follows from that of $\M^{\epsilon}$ and $\W_{ij}(s_{ij})$, the cut of the latter being only at $s_{ij}\geq (m_i+m_j)^2$.  

Crossing symmetry is manifest: the real axis at $s_{ij}\leq (m_i-m_j)^2$, corresponding to $i\ldots\to j\ldots$ kinematics, can be reached from the $ij^*\ldots\to\ldots$ channel at $s_{ij}\geq (m_i+m_j)^2$, involving antiparticle $j^*$,  by a path in the upper half-plane that avoids all cuts of $\W_{ij}(s_{ij})$.  
The sign flip from $\eta_i\eta_j$ in~\eqref{eq:WeinbergExpGR} is correctly reproduced, as well as the sign-preserving  combination $\eta_{i}\eta_{j} q_i q_j$ in \eqref{eq:WeinbergExpQED}, since crossing relates particles to antiparticles of opposite charge.  

The essential difference between these thresholds is that circling the physical one, $(m_i+m_j)^2$, crosses both square-root and logarithmic cuts in~\eqref{eq:AnalyticSoftQED} and \eqref{eq:AnalyticSoftGR}, leading to a discontinuity for $s_{ij}>(m_i+m_j)^2$, whereas circling the pseudothreshold $(m_i-m_j)^2$ crosses only square-root cuts, whose contributions cancel, yielding no discontinuity for $s_{ij}<(m_i-m_j)^2$.

The stripped amplitudes $\M_\E$ are also trivially hermitian analytic.

\item {\bf Regge scaling of the $2\to 2$ amplitude.}  

{\it Gravity.} A version of the Froissart-Martin bound \cite{Froissart:1961ux,Martin:1962rt}, originally proved for gapped theories, holds in  gravitational theories~ \cite{Haring:2022cyf} \footnote{It is interesting to contrast Regge scaling of the full amplitude with the property of single partial waves, which have been shown to violate polynomial boundedness for large 
$\E$ in the context of wave scattering off black holes \cite{Correia:2025enx}.}: it is not the amplitude that directly satisfies the bound, but the amplitude smeared in momentum transfer $q\in[\E,q_{\max}]$ with a suitable function~$\psi(q)$,  at~$\epsilon>0$. In particular, the smeared amplitude satisfies twice--subtracted dispersion relations, for $n\geq 0$
\begin{equation}\label{froisgrav}
\oint_{|s|\to\infty} \frac{ds}{s}
\int_\E^{q_{\max}} dq\, q^{2\epsilon}\, \psi(q)\, 
\frac{\M^{\epsilon}(s,-q^{2})}{s^{2+n}}
\;\longrightarrow\; 0 \, .
\end{equation}
 The proof relies on the eikonal expansion of the amplitude at sufficiently large and real values of~$s$~\cite{Haring:2022cyf}, see also \cite{Caron-Huot:2022ugt}. A result analogous to Eq.~\eqref{froisgrav}  can be obtained for the stripped amplitude $\M_{\E}$, defined in the limit $\epsilon\to0$, even without smearing in $q$,  because the eikonal regime can be extended outside the real axis---for large {\it complex} values of $s$:
\begin{equation}
\begin{split}\label{ampeikbla}
\int_{\E}^{q_{\max}} dq \psi(q) &\M^{\epsilon}(s,-q^2)\underset{|s|/q^2\to\infty}{\longrightarrow} \int_{\E}^{q_{\max}} dq\psi(q)\M^{\epsilon}(s,-q^2)|_{\text{eik}}\,,\\
&\M^{\epsilon}(s,t)|_{\text{eik}}=-\frac{8\pi Gs^2}{t}\exp\left[\frac{Gs}{\pi\epsilon} \left(\log s -\log( -s)\right)\left(\frac{-t}{\mu^2}\right)^{\epsilon}\, \right] \,.
\end{split}
\end{equation}
Indeed, the diagrammatic proof of leading eikonal amplitude,  which selects ladders and cross-ladder diagrams controlled by $G$,   does not rely on the reality of $s$. Under such an extension, the stripped amplitude has the asymptotic form,
\begin{equation}
\int_\E^{q_{\max}}\!\!\!\!dq\, \psi(q)\,
\M_{\E}(s,-q^{2})
\underset{|s|/q^2\to\infty}{\longrightarrow}
\int_\E^{q_{\max}}\!\!\!\!dq\, \psi(q)\,
\frac{8\pi G\, s^{2}}{q^{2}}
\exp\!\left[
\frac{G s}{\pi}\bigl(\log s - \log(-s)\bigr)
\log\!\frac{q^{2}}{\E^{2}}
\right] \,,
\end{equation}
where the divergent $\sim 1/\epsilon$ part has canceled between the soft factor \eqref{eq:AnalyticSoftQED} and the amplitude~\eqref{ampeikbla}.
Performing the $s$--integration over the large circle at infinity explicitly, as in Eq.~\eqref{froisgrav}, we find that it vanishes for odd $n$, while for even (and non-negative) $n$ it is proportional to $\pi-2\mathrm{Si}(G|s|\log(q^2/\E^2))+O(1/|s|)$. 
In fact the integral over the large circle decays as $1/|s|^{n+1}$. 
We conclude that an \emph{average Regge bound} also holds for the stripped amplitude: for $n\ge0$ it yields
% \bb{Performing explicitly the $s$--integration over the big circle at infinity, as in Eq.~\eqref{froisgrav}, we find that it vanishes for $n$ odd, and it is proportional to $\pi-2\mathrm{Si}(G|s|\log q^2/\E^2)+O(1/|s|)$. Therefore, the integral over the big circle decays as $1/|s|$.  In conclusion, an \emph{average Regge bound}  holds for the  stripped amplitude as well: for $n\geq 0$, it returns }
%Performing the integral explicitly over the big circle at infinity, we find that the Regge bound  holds for the (smeared) stripped amplitude as well: 
\begin{equation}
\oint_{|s|\to\infty} \frac{ds}{s}
\int_\E^{q_{\max}} dq\, \psi(q)\,
\frac{\M_{\E}(s,-q^{2})}{s^{2+n}}
\;\longrightarrow\; 0 \, .
\end{equation}

The same conclusion can be reached following similar steps to \cite{Haring:2022cyf,Caron-Huot:2022ugt} where one makes use of the smearing wavepacket $\psi(q)$ and of the eikonal formula for real kinematics. In our case, however, unitarity of the stripped amplitude is only approximate, see Section \ref{section:unitarity}, and therefore the Regge scaling is recovered---with this method, as opposed to the previous one---only in the scaling limit \eqref{eq:generalscaling}. 

{\it QED.} The Froissart--Martin bound for the full regulated $2\!\to\!2$ amplitude in QED
\begin{equation}
    \lim_{|s|\to+\infty} \M^{\epsilon}/s^2 =0\,,\qquad t\ \text{fixed} \,,
\end{equation}
is justified because very soft momenta are effectively cut off by working at finite $\epsilon>0$. This is obvious with a small photon mass as IR regulator, where we have $1/\epsilon=\log(\mu^2/m_\gamma^2)$ and $\M^\epsilon$ is the amplitude of a gapped theory, hence obeying the Froissart-Martin bound rigorously. 
Then, the asymptotic behavior of $\M^\epsilon_{\E}$ in QED, at $s\gg-t$ and fixed $t$, is determined entirely by the scaling of the exponential factor $\W$.  
In all cases of interest  $\W$ approaches a constant or further suppresses the amplitude at large $s$, ensuring that,
\begin{equation}
    \lim_{|s|\to+\infty} \M^{\epsilon}_{\E}/s^2 =0\,,\qquad t\ \text{fixed}\,.
\end{equation}
For example, in elastic $\pi^+\pi^0$ or  $\pi^{+}\pi^{-}$ scattering, one finds in this limit $\prod_{ij}\W^{\mathrm{QED}}_{ij}=\W^{\mathrm{QED}}(t)$, so Regge scaling is trivially preserved.    
The validity of this property should be verified case by case.

As for the Regge scaling of $\M_{\E}=\lim_{\epsilon\to 0^+} \M_{\E}^\epsilon$, that is whether the $\epsilon\to 0$ and $|s|\to \infty$ limits commute, we can not prove it in full generality. 
However, as the only gapless particles in this theory are spin-$1$, we expect no violation of the Froissart bound. Indeed, the leading contribution of gapless modes to the stripped amplitude in the eikonal regime $-t\ll |s|$, with QED-only insertions, scales as $\alpha s/t$, which is bounded by $s^2$.  
Supported by this intuition, we assume that for $n\geq 0$
\begin{equation}
\label{eq:ReggeQED}
\oint_{|s|\to \infty} \frac{ds}{s} \int_{\E}^{q_{\max}} dq \psi(q) \frac{\M_{\E}(s,-q^2)}{s^{2+n}} \to  0 \,,
\end{equation}
which allows us to derive at least twice--subtracted dispersion relations for $\M_{\E}$ in the next sections. 

Alternatively, as for the gravitational case, an approximate version of this  conclusion can be reached following the steps of \cite{Haring:2022cyf,Caron-Huot:2022ugt}, where the smearing wavepacket $\psi(q)$ allows to control the large partial--waves contribution.   

\end{enumerate}

\section{Unitarity of Stripped and Hard Amplitudes}
\label{section:unitarity}

The properties of the stripped amplitudes discussed above follow directly from those of the full amplitude and from the structure of the exponential factor.  
 We now need to establish an appropriate notion of unitarity for $\M_{\E}$.  
Stripped amplitudes depend effectively on two  couplings (in addition to other short-range interactions): $\alpha$ and $\alpha_\E=\alpha\log(M/\E)$ with electromagnetism (or $G M^2$ and $G_\E=G M^2\log(M/\E)$ in gravity) -- $M$ a hard scale. 
We are interested in the regime of \eqref{eq:generalscaling}, $4\pi \gtrsim \alpha_\E\gg\alpha$ (or $4\pi \gtrsim G_\E\gg GM^2$), and aim to control unitarity to all orders in $\alpha_\E$ (or $G_{\E}$) and any other short--distance coupling, while perturbatively in $\alpha$ (or $G$):
\begin{equation}
\begin{split}
\mathcal{M}_{\E}(\alpha_\E,\alpha)&= \mathcal{M}^{(0)}(\alpha_\E)+\alpha\, \mathcal{M}^{(1)}(\alpha_\E)+\ldots\,,  \\ 
\mathcal{M}_{\E}(G_\E,G)&= \mathcal{M}^{(0)}(G_\E)+G E^2\, \mathcal{M}^{(1)}(G_\E)+\ldots\,.
\end{split}
\end{equation}
 In particular, we focus on the properties of $\mathcal{M}^{(0)}(\alpha_\E)$ and $\mathcal{M}^{(0)}(G_\E)$. 
Crucially, these  are  linked to  hard amplitudes by the same limit:
\begin{equation}
\label{eq:m0ismhard}
%\begin{split}
  \mathcal{M}^{(0)}(\alpha_\E)\,\,\,= 
\lim_{\substack{\alpha\to 0 \\ \E \to 0 \\ \alpha\log(M/\E)\to \alpha_{\E}}}
    \M_{\E}(\alpha_\E,\alpha)\,\,\,\,
=
\lim_{\substack{\alpha\to 0 \\ \E \to 0 \\ \alpha\log(M/\E)\to \alpha_{\E}}} 
 \M^{\mathrm{hard}}_{\E}(\alpha_\E,\alpha)\,, \qquad 
\end{equation}
and analogously for gravity.
This implies that the leading contribution to $\M_{\E}$ in \eqref{eq:hardsubleadingseparation} originates from the hard amplitude,  whose unitarity we are able to  control explicitly.  Subleading soft contributions $\delta_{\mathrm{sub.\,soft}}$ in \eqref{eq:hardsubleadingseparation}, associated with the  IR-finite part of a soft-loop, decouple, since they
are suppressed by $\alpha$ (or $G$), rather than by $\alpha_\E$ (or $G_\E$).

\subsection{Unitarity of Hard Amplitudes}
\label{subsection:unitarity}

We now establish unitarity of \(\M^{\mathrm{hard}}_{\E}\)  (at all orders in the couplings), which in turn implies unitarity of \(\M^{(0)}(\alpha_\E)\) (at leading order in $\alpha$).  
Intuitively,  \(\M^{\mathrm{hard}}_{\E}\) is expected to be unitarity because  QED with a massive photon with $m_{\gamma} = \E$ is unitary. In the scaling limit \eqref{eq:m0ismhard}, $\M_{\E}^{\mathrm{hard}}$ coincides with the amplitudes for this theory, once the massive photon longitudinal polarizations are suitably decoupled by adjusting the vertices in the theory. In what follows,
we formalize this intuitive picture in a way that holds also for gravity. 

Our strategy is to construct an explicit Hermitian Hamiltonian $H^{\mathrm{hard}}_{\E}$ acting on a positive-definite Hilbert space, with in--to--out matrix elements that reproduce the hard amplitudes. Once such a Hamiltonian is proven to exist, unitarity of $\M^{\mathrm{hard}}_{\E}$ follows.

We begin with a tree-level ansatz for $H^{\mathrm{hard}}_{\E}$, and then correct it order by order in perturbation theory. 
Our starting point is  the gauge-fixed Hamiltonian of the full, regulated and renormalized theory, truncated so that only physical (positive-norm) modes remain, for all $\boldsymbol{k}$. We then restrict the Hilbert space to the subset of these physical modes with $|\boldsymbol{k}|>\E$ and include momentarily no additional vertices or kinetic terms.
We illustrate this in QED in Coulomb gauge,\footnote{One could also work \textit{e.g.} in Lorenz gauge (together with a subsidiary Gupta–Bleuler condition that selects physical states), even though Lorentz symmetry is still explicitly broken in $H_{\E}^{\text{hard}}$ by $O(\E)$ effects.
} where the counting of physical degrees of freedom is manifest:
\begin{equation}
\label{eq:HamiltonianCoulombGauge}
H^{\mathrm{hard}}_{\E}=\sum_{h=\pm}\int_{|\boldsymbol{k}|>\E}\frac{d^3\boldsymbol{k}}{(2\pi)^3}\left\{
\omega(\boldsymbol{k})a_{h}^{\dagger}(\boldsymbol{k})a_{h}(\boldsymbol{k})-e\left[a_{h}(\boldsymbol{k})\boldsymbol{\varepsilon}^{h}(\boldsymbol{k})\cdot\boldsymbol{J}+\mathrm{h.c.}\right] +V_{\mathrm{Coul}}
\right\}+c.t.+\ldots \,,
\end{equation}
with $\boldsymbol{\varepsilon}^{\pm}(\boldsymbol{k})$ are transverse polarizations, $V_{\mathrm{Coul}}$  the instantaneous Coulomb potential, $\boldsymbol{J}$ the matter current, and dots stand for all other terms which do not involve photons.
The states are relativistic, with $\omega(\boldsymbol{k})=|\boldsymbol{k}|$, and the counterterms {\it c.t.} are the same ones as  in the full theory. 
At tree level this Hamiltonian is Hermitian, preserves the space of hard modes, and generates tree-level hard amplitudes among positive-norm states.

When radiative effects are included, the corrections to the observables are UV-finite, since the UV divergences are identical to those of the parent theory and are thus absorbed by the counterterms. However, the absence of soft modes in loops can introduce additional {\it finite} corrections---mass shifts, velocity renormalization, and Lorentz-breaking vertices---which do not exist in the full theory.\footnote{No non--local finite correction is generated, since no soft mode is contained in the hard hamiltonian. Non--local corrections to the quantum action come from loops of hard modes, perfectly consistent with unitarity. }

These finite corrections are benign: they can be cancelled by finite counterterms—merely a harmless scheme choice. That is, we amend the original ansatz by adding minus the finite counterterms responsible for mass shifts, velocity corrections, Lorentz-breaking vertices, and so on.
While this may appear to be an ad-hoc fine-tuning---albeit a legitimate one---it is actually natural, since soft modes exist and contribute through $\delta_{\mathrm{sub. soft}}$ in \eqref{eq:hardsubleadingseparation} to restore the Lorentz invariance of $\M_\E$, to which $\M^{\mathrm{hard}}_{\E}$ converges.
Changing the {\it c.t.} by a finite amount---which is actually subleading with respect to the leading logarithmic effects\footnote{This is because all $\log\E$–enhanced terms from the soft region have been removed by the exponential $\W$.}---is effectively equivalent to shifting some finite terms between $\delta_{\mathrm{sub. soft}}$ and $\M^{\mathrm{hard}}_{\E}$ in \eqref{eq:hardsubleadingseparation}.  Adopting this prescription for $H^{\mathrm{hard}}_{\E}$, the associated hard amplitude $\M^{\mathrm{hard}}_{\E}$ propagates only positive-norm modes with relativistic dispersion and $|\boldsymbol{k}|>\E$. 

Everything said for QED carries over to gravity: one may start from a fully gauge-fixed Hamiltonian where only the physical helicity modes  $h=\pm2$  propagate with positive norm, and then possibly adjust finite parts in the counterterms.\\ 

Once a Hermitian Hamiltonian acting on a positive-definite space is constructed, the generalized optical theorem follows immediately from $S^{\mathrm{hard}\,\dagger}_{\E}S^{\mathrm{hard}}_{\E}=\mathbb{I}$ and $S^{\mathrm{hard}}_{\E}=\mathbb{I}+iT$~\footnote{The hard hamiltonian corresponds to short--range interactions with partial waves decaying as $\ell^{-3/2}$ times an oscillatory factor that helps convergence, allowing the splitting between identity and interactions.},  where the identity is restricted to hard photons and gravitons, but otherwise acts as usual on the rest of the states.  In particular, for a $2\to2$ hard amplitude, the generalized optical theorem is  
\begin{equation} \label{eq:opticaltheorem}
 2\,\overline{\text{Im}}\,\M^{\mathrm{hard}}_{\E}(a\,b \to c\,d)=\sum_{X}\M^{\mathrm{hard}}_{\E}(c\,d\to X)^{*}\M^{\mathrm{hard}}_{\E}(a\,b \to X) \,,
\end{equation}
where the sum over intermediate states $X$ excludes soft photons and gravitons, and we introduced a generalized imaginary part~\cite{Caron-Huot:2022ugt},
\begin{equation}
    \overline{\text{Im}}\,\M^{\mathrm{hard}}_{\E}(a\,b \to c\,d)\equiv\M^{\mathrm{hard}}_{\E}(a\,b \to c\,d)- \M^{\mathrm{hard}}_{\E}(c\,d \to a\,b)^*\,.
\end{equation}
We stress that the unitarity of hard amplitudes is an exact statement, and does not rely on the scaling limit  \eqref{eq:generalscaling} -- it is only the connection with $\M_\E$ that does.

The previous arguments are at the operator level, relying on $H^{\mathrm{hard}}_{\E}$. They can as well  be recasted diagrammatically, in a manifest form, via cutting rules. The derivation from the largest-time equation~\cite{Veltman:1963th} remains indeed valid in the presence of $\E$;  one can \textit{e.g.} follow closely the treatment of~\cite{Anselmi:2016fid} with an appropriate gauge fixing.  
The only modification in the cutting rules is the replacement of the propagator $\Delta(k)$ by $\theta(|\boldsymbol{k}|-\E)\Delta(k)$, which does not alter causality, energy scaling, or the locality of interactions.\\

Finally, from equations~\eqref{eq:m0ismhard} and~\eqref{eq:opticaltheorem}, it follows:    
\begin{enumerate}[(e)]
\item {\bf Positivity and unitarity}
\begin{equation} \label{eq:opticaltheoremStripped} 
  2\,\overline{\text{Im}}\,\M^{(0)}(a\,b \to c\,d) %+O(\alpha,\E)
= \lim_{\substack{\alpha\to 0 \\ \E \to 0 \\ \alpha\log(M/\E)\to \alpha_{\E}}}\sum_{X}\M^{\mathrm{hard}}_{\E}(c\,d\to X)^{*}\M^{\mathrm{hard}}_{\E}(a\,b \to X) \succeq 0\,. 
\end{equation}
\end{enumerate}
An analogous statement holds in gravity in the limit where $G_\E$ is held fixed.  \\

\subsection{Functionals: Unitarity and $1/t$--pole}
\label{subsubsect:functionals}

There are instances where unitarity of $\M^{(0)}_\E$ is not sufficient, because the term $\M^{(1)}_\E$ is enhanced by the specific observable being considered.
The paradigmatic example of this effect is the integration of a $\alpha/t$ pole in the $2\to 2$ amplitude. Even in the hard region of phase space, $\int_{\E}^M dt\, \alpha/t = \alpha \log(M/\E)$, so that an $O(\alpha)$ term in $\M_\E^{\mathrm{hard}}$ effectively becomes an $O(\alpha_\E)$ term in the integral. This mismatch arises from the propagation of long-range mediators in the $t$-channel and can be anticipated and computed in the EFT.

To remedy this, we can instead study functionals of the stripped amplitude, $\mathcal{F}[\M_{\E}]$, representing physical observables, rather than the amplitude itself.  
Examples include dispersive integrals, as well as integrated cross sections.  
These functionals too admit  an expansion in $\alpha$ at finite $\alpha_\E$:
\begin{equation}
\label{eq:exandingfunctionals}
\mathcal{F}[\mathcal{M}_{\E}](\alpha_\E,\alpha)
=\mathcal{F}[\mathcal{M}_{\E}](\alpha_\E,0) 
+ \alpha\, \mathcal{F}^{(1)}[\mathcal{M}_{\E}](\alpha_\E)+\ldots
\end{equation}
(analogously in $G$), with  the leading contribution  in the regime \eqref{eq:generalscaling}  captured by $\mathcal{F}[\mathcal{M}_{\E}](\alpha_\E,0)$. 
Importantly, the functional and the limit do not necessarily commute, 
\begin{equation}\label{mismatch}
\lim_{\substack{\alpha\to 0 \\ \E \to 0 \\ \alpha\log(M/\E)\to \alpha_{\E}}}  \mathcal{F}[\mathcal{M}_{\E}](\alpha_\E,\alpha)\neq \mathcal{F}[\mathcal{M}^{(0)}(\alpha_\E)]\, ,
\end{equation}
forcing us to extend the  arguments of the previous section to a particular class of functionals.
 
Anticipating the discussion of the following sections, the family of functionals we are interested in is~\footnote{In fact, we will further integrate these functionals in $s$, but since the integrals are convergent and can be exchanged with the scaling limit, the $s$--integration is not relevant for this present discussion.}
\begin{equation}
\label{eq:classfunctional}
        \int_{\E}^{q_{\max}}\!\!dq \psi(q)\,\overline{\text{Im}}\,\M_{\E}(s,-q^2)\,,
\end{equation}
for which the following property holds: 
\begin{enumerate}[(e')]
\item {\bf Functional unitarity}
\begin{equation} \label{eq:functionalopticaltheorem} 
  \lim_{\substack{\alpha\to 0 \\ \E \to 0 \\ \alpha\log(M/\E)\to \alpha_{\E}}} \int_{\E}^{q_{\max}} \!\!dq~ \psi(q)\,\overline{\text{Im}}\,\M_{\E}(s,-q^2)=\!\!\!\!\!\lim_{\substack{\alpha\to 0 \\ \E \to 0 \\ \alpha\log(M/\E)\to \alpha_{\E}}}\frac{1}{2}\int_{\E}^{q_{\max}}\!\!dq~ \psi(q)\,\M^{\mathrm{hard}\,\dagger}_{\E} \M^{\mathrm{hard}}_{\E} \,,
\end{equation}
\end{enumerate}
where $\psi(q)$ is a suitable smearing function, such that $\psi(q)\to q$ as $q\to0$, and $s\gg q^2_{\max}$. An analogous statement holds in gravity, in the limit where $G_\E$ is held fixed. 

To prove \eqref{eq:functionalopticaltheorem}, it is enough to address the mismatch \eqref{mismatch} between amplitude and functional unitarity. The mismatch is entirely due to photon (graviton) exchanges generating terms of the form $\alpha/t$ ($G/t$) that appear subleading in $\alpha$ (or $G$) at the amplitude level. Such terms are always there with gravity, but are absent in QED if the scattering involves neutral particles, like $\pi^+\pi^0\to \pi^+\pi^0$.

At tree-level ---order $O(\alpha)$ or order $O(G)$---Eq. \eqref{eq:functionalopticaltheorem} is manifestly valid, since the $1/t$ contribution to $\M_{\E}$ originates from the hard photon (graviton) exchange. At loop--level, the full amplitude receives insertions of hard, leading soft, and subleading soft photons (or gravitons). The subleading soft insertions clearly provide further suppression by powers of $\alpha$ or $G$, and drop out after taking the scaling limit, whereas the leading soft modes are factored out by the Weinberg exponential in the definition of $\M_{\E}$. Therefore, we conclude that any left--over correction relevant for functional unitarity of $\M_{\E}$ is controlled by $\M^{\mathrm{hard}}_{\E}$.

This picture can be made systematic by hunting down $t$--channel singularities of $\M_{\E}$ via the eikonal expansion. For the scattering of charged equal-mass particles, the leading term in this expansion is
\begin{equation}\label{eq:imeikonalQED}
    \overline{\text{Im}}\,\M_{\E}(s,t)|_{\text{eik}}=q_iq_j\frac{8\pi\alpha s}{t}\sin\left(\frac{\alpha}{\beta_s}\log\frac{-t}{\E^2}\right) \,,
\end{equation}
with $\beta_s\equiv\sqrt{s(4m^2-s)}/(s-2m^2)$, while in the presence of gravity (considering neutral massless particles for simplicity)
\begin{equation}\label{eq:imeikonalgrav}
    \overline{\text{Im}}\,\M_{\E}(s,t)|_{\text{eik}}=-\frac{8\pi Gs}{t}\sin\left(Gs\log\frac{-t}{\E^2}\right) \,.
\end{equation}
For $-t\gg\E^2$---but still much smaller than any other scale--- it is straightforward to verify that Eqs. \eqref{eq:imeikonalQED} and \eqref{eq:imeikonalgrav} match the corresponding leading eikonal expansion of $\overline{\text{Im}}\,\M^{\mathrm{hard}}_{\E}(s,t)|_{\text{eik}}$. As for the region $-t\sim\E^2$, we observe that $\overline{\text{Im}}\,\M_{\E}(s,t)|_{\text{eik}}$ is dominated by the tree-level exchange, which again originates from $\M^{\mathrm{hard}}_{\E}$.
Incidentally, the $1/t$ singularities in \eqref{eq:imeikonalQED} and \eqref{eq:imeikonalgrav} provide clear examples of cases in which the exchange of the limit with the functional is not allowed.

In conclusion, property (e) unitarity \eqref{eq:opticaltheoremStripped} and (e') functional unitarity \eqref{eq:functionalopticaltheorem} allow us to derive positivity bounds in the presence of long-range interactions. We specifically show positivity properties of the partial--wave decomposition in Section~\ref{sect:dispersion_relations} and use them to show that bounds exist. We formalize them in Section~\ref{Sec:BoundsEM} and Section~\ref{sec:positivitygravity}.

Before ending this subsection it is perhaps worth stressing that other approximation schemes may apply under extra assumptions. An obvious example is when the stripped amplitude $\M_{\E}$ is assumed to be well approximated by a meromorphic, tree--level amplitude even \emph{in the UV}. This is similar to the limit that is considered in the bootstrap program of large--N gauge theories~\cite{Albert:2022oes,Fernandez:2022kzi,Berman:2023jys} or in certain tree--level stringy UV completion of Einstein--gravity, e.g. \cite{Camanho:2014apa}. In these special cases, even the orders $O(\alpha)$ and $O(G)$ are under control and manifestly  unitary, $\M_{\E}\big|_{\mathrm{tree}}=\M^{\epsilon}\big|_{\mathrm{tree}}=\M^{\mathrm{hard}}_{\E}\big|_{\mathrm{tree}}$ for $\E/M\ll1$,  giving rise to $\int_{\E}^{q_{\max}} \!\!dq~ \psi(q)\,\overline{\text{Im}}\,\M_{\E}(s,-q^2) \succeq 0$ up to power corrections $O(\E/M)$ and IR--finite loops. The accuracy is to order $O(G,\alpha)$ included, with error starting at $O(G^2,\alpha^2, g_*^2\alpha, g_*^2G, \alpha \alpha_{\E},\ldots)$. 
Importantly, as opposed to earliest studies involving $\M^\epsilon\big|_{\mathrm{tree}}$, the corrections to $\M_{\E}\big|_{\mathrm{tree}}$ are  IR--finite and can be small in these class of theories.

\subsection{Collinear enhancements}

In addition to soft divergences, scattering massless charged particles also leads to collinear IR divergences. When the charged particles carry a finite mass (as we assume here), the collinear singularity is regulated by that mass. In our setup, therefore, no genuine collinear divergence is present. 

At sufficiently high energies, however, collinear effects may re--emerge in the form of finite but large logarithms of the type $\alpha \log(E/m)$ in QED. There are no collinear divergences instead in gravity, see \textit{e.g.} \cite{Beneke:2022pue,Akhoury:2011kq}. 
One may thus wonder whether the $O(\alpha)$ contributions in QED that we have systematically discarded in order to prove unitarity could, after being enhanced by such collinear logarithms, compete with the $\alpha_{\E}$ terms we have instead retained. 

This does not occur. In any theory with non--trivial soft IR divergences, the soft and collinear regions always (partly) overlap, so that the $O(\alpha_{\E})$ contributions are themselves enhanced by the same collinear logarithms. This is manifest in the explicit form of the Weinberg exponential $\prod_{ij}\W^{\text{QED}}_{ij}$ in \eqref{eq:AnalyticSoftQED}.  As a result, the leading contributions to the stripped amplitudes are governed by $\alpha_{\E}$ at every order in perturbation theory and the terms proportional to $\alpha$ remain subleading.

\subsection{Connection to Faddeev-Kulish Amplitudes}
\label{subsect:Faddeev-Kulish}

In the scaling limit \eqref{eq:generalscaling} all IR-finite amplitudes $\M^{FK}_{\E}$ defined \textit{\`a la} Faddeev-Kulish \cite{Kulish:1970ut,Kapec:2017tkm,Hannesdottir:2019opa} match the stripped amplitude $\M_{\E}$ by construction. Indeed the leading-order effect of the modified asymptotic dynamics of Faddeev--Kulish is exactly to remove the virtual\footnote{As for additional real soft photons, their contribution vanishes in FK amplitudes in the limit \eqref{eq:generalscaling}.} soft photon factors, namely
\begin{equation}
    \lim_{\substack{\alpha\to 0 \\ \E \to 0 \\ \alpha\log(M/\E)\to \alpha_{\E}}}\M^{FK}_{\E}=\lim_{\substack{\alpha\to 0 \\ \E \to 0 \\ \alpha\log(M/\E)\to \alpha_{\E}}}\lim_{\epsilon\to0^+}\frac{\M^\epsilon}{\prod_{ij}\W^{\mathrm{QED}}_{ij}(s_{ij})}=\lim_{\substack{\alpha\to 0 \\ \E \to 0 \\ \alpha\log(M/\E)\to \alpha_{\E}}}\M_{\E}=  \M^{(0)}(\alpha_\E) \,.
\end{equation}
This can be shown by repeating Weinberg's computation and taking into account soft photons from the dressing operators. A scale $\E$ is inevitably introduced in the construction of IR-finite amplitudes, independently of the specific modeling of the asymptotic dynamics, either as an hard cut-off or as a renormalization-like scale \cite{Hannesdottir:2019opa}. For instance, a possible choice---among infinitely many--- for the asymptotic hamiltonian is the soft--hamiltonian itself, $H-H^{\mathrm{hard}}_{\E}$. 

Thus, all the properties such as analyticity and unitarity apply to Faddeev-Kulish amplitudes as well, in the limit \eqref{eq:generalscaling}. The converse is also interesting: manifest unitarity of FK amplitudes implies unitarity of $\M^{(0)}(\alpha_\E)$. 

In this sense, the leading contribution of $\M_{\E}$, isolated through the limit \eqref{eq:generalscaling}, captures the universal leading contribution to the scattering. For generic exclusive quantities, even when IR finite, corrections will instead be sensitive to the very details of the asymptotic dynamics, as it is evident from the ambiguities in Faddeev-Kulish-like amplitudes. This sensitivity disappears when considering sufficiently inclusive observables\footnote{In fact, as discussed in \cite{Carney:2018ygh}, interference effects are IR sensitive even for inclusive observables, leading to decoherence unless the states are properly dressed.} \cite{Forde:2003jt,Hannesdottir:2019opa,Agarwal:2021ais} or energy--energy correlators \cite{Moult:2025nhu}.

\section{Dispersion Relations}
\label{sect:dispersion_relations}

Stripped amplitudes $\M_{\E}$  are IR-finite, analytic,  exhibit Regge scaling, and satisfy a form of unitarity \eqref{eq:functionalopticaltheorem}.
These properties allow us to derive twice-subtracted dispersion relations for $\M_{\E}$, establishing UV/IR connections at leading order in $\alpha$ or $G$ (hence free from IR divergences), but at all orders in $\alpha_\E$ and $G_\E$.

Given  an amplitude for the elastic $2\to 2 $ scattering of spin-0 particles  and the associated stripped amplitude $\M_{\E}$, as well as a smearing function $\psi$ with $\psi(q)\to q$ for small $q$,
we define the smeared arcs,
\begin{equation}
\label{eq:IRArcstripped}
    \A_n[\psi,\M_\E]
    \equiv
    \oint_{C} \frac{dz}{2\pi i}\,
   \int_{\E}^{q_{\max}} \! dq\, \psi(q) \mathcal{K}_n(z,q){\M_{\E}(z,q)}  \,,
\end{equation}
as a complex/real integral in the variables $z(s,t)$ and $q(s,t)$.
The kernel $\mathcal{K}_n$ satisfies $\mathcal{K}_n\to 1/s^{n+3}$ as $s\to \infty$ while having poles of total order $n+3$ within the contour~$C$. The variable $q(s,t)$ is such that $q^2\sim-t$ for $-t\ll s$. The definition \eqref{eq:IRArcstripped} is sufficiently vague as to accommodate different types of dispersion relations that appear in the literature. For instance, in fixed-$t$  dispersion relations we have $z=s$ and the smeared amplitude  is simply $\M_{\E}(z,q)=\M_{\E}(s,t)$ with,
\begin{equation}\label{fixedtstuff}
   q^2=-t\quad \textrm{fixed},\quad\quad\quad \mathcal{K}_n=\frac{1}{(s-s_0)^{3+n}}\,.
\end{equation}

On the other hand, for amplitudes that are $t\leftrightarrow u$ symmetric (like $\pi^+ \pi^+\to \pi^+ \pi^+$), or $s-t-u$ symmetric (like $\pi^0 \pi^0\to \pi^0 \pi^0$), the definition \eqref{eq:IRArcstripped} might equally well describe  crossing symmetric dispersion relations \cite{Bellazzini:2025shd,Auberson:1972prg,Sinha:2020win,Li:2023qzs}. In the former case, for instance, we identify ${\M}_{\E}(s,q)$ with $ \M_{\E}(s,t=t_{+}(s,q))$, where,
\begin{equation}
    q^2=\frac{tu}{s}\,,\quad\quad \,\quad\quad t_{\pm}(s,q)\equiv-\frac{s-4m^2}{2}\left(1\mp\sqrt{1-\frac{4q^2s}{(s-4m^2)^2}}\right)\,. %\qquad -s-t_{\pm}+4m_{\pm}^2=t_{\mp}\,
\end{equation}
with $m$ the pion mass. We review all these cases in detail in  Appendix~\ref{appendix:TUSymm}, where we provide the corresponding kernels $\mathcal{K}_n$, contours $C$, and boundaries of integration $q_{max}$. The generality of \eqref{eq:IRArcstripped} allows us to transcend from the nitty gritty of dispersion relations and appreciate the universal advantages provided by stripped amplitudes.

UV/IR relations follow from identifying a scale $M$ that separates the IR $|s|\leq M^2$, where the amplitude is computable in an EFT in terms of its Wilson coefficients and coupling constants, and the UV $|s|> M^2$ with unknown unitary and causal physics.\footnote{The UV/IR terminology for arcs can be slightly misleading, since $|t|<M^2$, meaning that the momentum exchange always lies in the IR. Only $s$ probes UV regions. For instance, in gravity, at $s\gg 1/G$ and fixed $t$, the amplitude is IR-calculable through the eikonal expansion~\cite{Haring:2022cyf,Bellazzini:2022wzv}.} The contour $s\in C$ lies entirely in the IR. We can then provide a UV representation of arcs by deforming   $C$   around the branch cuts  associated with the physical regions of the amplitude in the relevant channels  $k=s,u,t$,  into a large circle at infinity. The latter vanishes thanks to Regge scaling \eqref{eq:ReggeQED}, and we obtain, using crossing symmetry,
\begin{eqnarray}
\label{eq:integraldiscgeneric0}
        \A_n[\psi,\M_{\E}]=\sum_{\mathrm{cuts}\,k}\int_{M_k^2}^{+\infty}\frac{ds}{2\pi i} \int_{\E}^{q_{\max}} dq \,\psi(q){(\text{Disc}_{k}[\M^k_{\E}}]\mathcal{K}^k_n)(s,q)\,,
\end{eqnarray}
where, for compactness, we have changed variables in order to group all branch-cuts under the same $s$-integral over the discontinuity of the $k$-channel amplitude $\M^k_{\E}$, with the corresponding kernel $\mathcal{K}^k_n$, starting at  $M_k^2$. The EFT cut-off can be interpreted as $M=\textrm{min}_k M_k$.
Hermitian analyticity of the stripped amplitude implies that  $\text{Disc}\M_\E=\M_\E-\M_{\E}^{\dagger}=2i \overline{\text{Im}}\M_\E$, which we evaluate in terms of the physical amplitude in that channel, $\M_{\E}^{k}$.\footnote{By the ``physical amplitude'' in a particular channel, we mean one with physical scattering angle and center-of-mass energy.} For instance, the fixed-$t$ version of this generic expression reads,
\begin{equation}
\label{eq:integraldiscgeneric}
        \A_n[\psi,\M_{\E}]=\int_{M^2}^{+\infty}\frac{ds}{\pi } \int_{\E}^{q_{\max}} dq \psi(q)\left[\frac{\overline{\text{Im}}\,\M_{\E}^s(s,-q^2)}{(s-s_0)^{3+n}}+\frac{(-1)^n \overline{\text{Im}}\,{\M}^u_{\E}(s,-q^2)}{(s-q^2-4 m^2+s_0)^{3+n}}\right] \,,
\end{equation}
with ${\M}^u_{\E}(s,t)$  the stripped amplitude for the $s$--$u$ crossed process, and $m^2\equiv (m_i^2+m_j^2)/2$  the average  squared external mass for  $ij$ scattering, stemming from the change of variables to the $u$-channel discontinuity. Expressions for $su$ and $stu$ crossing-symmetric dispersion relations are found in Appendix~\ref{appendix:TUSymm}.\\

We first focus on theories with electromagnetic interactions: here, arcs of the stripped amplitudes are IR-finite functions of the couplings $\alpha_\E$ and $\alpha$, as well as all Wilson coefficients in the EFT. In line with the discussion in Section \ref{subsubsect:functionals} we expand in the smallest coupling:
\begin{equation}
\label{eq:expandingfunctionals}
\A_n[\psi,\M_{\E}]( \alpha_\E,\alpha)= \A^{(0)}_n[\psi,\M_{\E}]( \alpha_\E)+ \alpha\, \A^{(1)}_n[\psi,\M_{\E}]( \alpha_\E)+\ldots 
\end{equation}
while keeping $\alpha_\E$ (and possibly also the Wilson coefficients) finite.   
The leading $\A^{(0)}_n$ term is constrained by (functional) unitarity Eq.~\eqref{eq:functionalopticaltheorem}, yielding 
\begin{eqnarray}
\label{eq:integraldiscgeneric0}
        \A^{(0)}_n[\psi,\M_{\E}]=\lim_{\substack{\alpha\to 0 \\ \E \to 0 \\ \alpha\log(M/\E)\to \alpha_{\E}}}\sum_{\mathrm{cuts}\,k}\int_{M_k^2}^{+\infty}\frac{ds}{\pi} \int_{\E}^{q_{\max}} dq \,\psi(q){(\overline{\text{Im}}_{k}[\M^{k,\text{hard}}_{\E}}]\mathcal{K}^k_n)(s,q)\,,
\end{eqnarray}
thereby leading to the positivity properties of arcs.

Positivity is better emphasized by exploiting rotational invariance to expand the hard amplitudes of channel-$k$ into Legendre polynomials $P_{\ell}$, with real coefficients $a^k_{\E,\ell}$. Doing so, we obtain,
\begin{equation}
\label{eq:UVrepresentation0}
        \A^{(0)}_n[\psi,\M_{\E}]=\!\! \!\!\!\!\lim_{\substack{\alpha\to 0 \\ \E \to 0 \\ \alpha\log(M/\E)\to \alpha_{\E}}} \!\!\!\!  
        \sum_{\mathrm{cuts}\,k}
        \SumInt_{\,\,\,\,\,\,\,M_s^2}^{\,\,\,\,\,\,\,\,\,\,\,\,\,\,\infty}\frac{ds}{2\pi s^{3+n}} P_{\ell,n}^{\psi,k}(s)|a^k_{\E,\ell}|^2(s) 
\end{equation}
where we have introduced the notation $\SumInt_{a}^b=\sum_{\ell}(2\ell+1)\int_a^b$
 and defined,
\begin{equation}\label{eqvagaevuota}
P_{\ell,n}^{\psi,k}(s)\equiv  \int_{0}^{q_{\max}} \!\!\!\!\!dq\,\psi(q) P_{\ell}(\cos\theta^k(s,q))\mathcal{K}^k_n=
\int_{\E}^{q_{\max}} \!\!\!\!\!dq\,\psi(q) P_{\ell}(\cos\theta^k(s,q))\mathcal{K}^k_n+O\left(\alpha,{\E^2}\right)\,,
\end{equation} 
with $\theta^k$ the physical angle in channel $k$. 

For reasons that will become apparent in \eqref{eq:positivefunctionals}, we have extended the
$q$–integration in \eqref{eqvagaevuota} down to $q=0$, using the fact that hard amplitudes do not propagate photons
(or gravitons) with $q<\E$. Crucially, the resulting mismatch,
\begin{equation}
\label{eq:mismatchInt}
\int_0^\E dq\, \psi(q)\,\mathcal{K}^k_n\,\overline{\mathrm{Im}}\,\M_{\E}^{k,\mathrm{hard}}
= O\!\left(\alpha,\E\right)\,,
\end{equation}
is not $\log \E$–enhanced and disappears in the limit in
\eqref{eq:generalscaling} that extracts $\A^{(0)}$.\footnote{By contrast, we do not
extend the integration in the IR representation of the arc
\eqref{eq:IRArcstripped} to $q\to 0$.}

Indeed, as $\E\to 0$, the dominant contributions to this integral arise from the (would--be)
 $q^2\sim 0$ singularities of Feynman diagrams contributing to
$\M_{\E}^{\mathrm{hard}}$, which are better behaved  than those of
$\M_{\E}$ and $\M^{\epsilon}$ itself. As sanity check, let us  consider the integral of the imaginary part of the box
diagram, whose most relevant contribution in this regime is approximated by
\begin{equation}
\!\!\!\!\!\int_0^\E q\,dq \!\!\adjustbox{valign=c,scale={1}{1}}{
\includegraphics[scale=0.3]{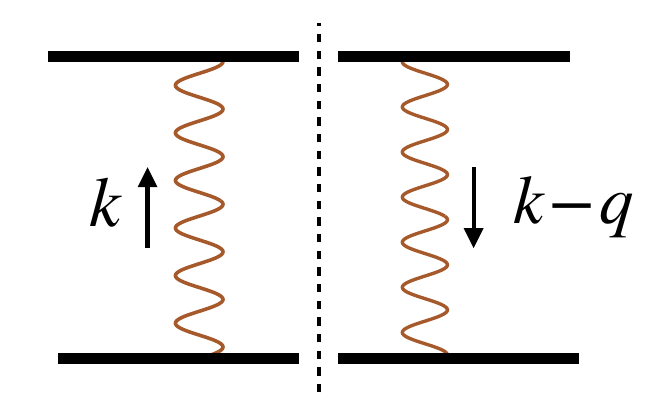}
    }\!\!\! \sim \int_0^\E q\,dq\left[ \frac{(8\pi\alpha s )^2}{4s}\int \frac{k\,dk}{2\pi}\frac{\theta(k-\E)}{k^2}\frac{\theta(|k-q|-\E)}{(k-q)^2}\right]\sim 2\pi \alpha^2 s\,.
\end{equation}The hard theta-functions remove the potentially divergent regions $k=0$ and $k=q$ so that the remaining domain satisfies $q\ll k$ and the integrand can be safely expanded in this limit. This contribution is independent of $\E$ and  scales as $O(\alpha^2)$ for
$\M_{\E}^{\mathrm{hard}}$, in contrast to the $O(\alpha_{\E}^2)$ behavior of
$\M_{\E}$ (and $O(\alpha^2/\epsilon^2)$ in $\M^{\epsilon}$). More generally, an eikonal analysis of the most singular contributions, given by the large partial-waves starting from $l_*\ll\sqrt{s}/\E$,
shows that the mismatch integral in \eqref{eq:mismatchInt} is subleading in the scaling
limit~\eqref{eq:generalscaling} and regular as $\alpha\to0$.\\

For fixed-$t$ dispersion relations, the amplitude has branch cuts in the $s\in \mathbb{C}$ plane along the $k=s,u$ channels, and there \eqref{eqvagaevuota} reads:
\begin{equation}
\begin{split}
P_{\ell,n}^{\psi,s}(s)\equiv & \int_{0}^{q_{\max}} dq\,\psi(q) P_{\ell}\left(1+2\frac{q^2}{s-4m^2}\right) \frac{s^{3+n}}{(s-s_0)^{3+n}}\,,\\
{P}_{\ell,n}^{\psi,u}(s)\equiv & \int_{0}^{q_{\max}} dq\,\psi(q) P_{\ell}\left(1+2\frac{q^2}{s-4m^2}\right)\frac{(-1)^ns^{3+n}}{(s+t-4 m^2+s_0)^{3+n}} \,.
\end{split}
\end{equation} 
For $tu$-- and $stu$--symmetric dispersion relations, the analogous expressions of these $P_{\ell,n}^{\psi,s}(s)$ are reported in Eq. \eqref{eq:tuPelln} and \eqref{eq:stuPelln}.

The UV representation \eqref{eq:UVrepresentation0} makes manifest that the leading contribution to  arcs $\A^{(0)}_n[\psi,\M_{\E}]$ is a sum of  {unknown} but \emph{positive} terms $|a^k_{\E,\ell}|^{2}(s)$, modulated by \textit{known} functions whose degree of negativity is controlled by the $P^{\psi,k}_{\ell,n}$.  Since Legendre polynomials are bounded both above and below, the amount of negativity can in turn be bounded for a given choice of $\psi$.

Following \cite{Caron-Huot:2021rmr}, bounds on EFT couplings can thus be derived by identifying functionals $\psi(q)$ such that we get positive contributions, 
\begin{equation}
\label{eq:positivefunctionals}
P^{\psi,k}_{\ell,n}(s) \geq 0 \qquad \forall \ell\in\mathbb{N}\,,\,\,  \forall s\geq M^2\, ,
\end{equation}
for all relevant channels $k$.
Such $\psi$ enforce positivity of the IR representation
\begin{align}\label{eq:zeroordpos}
\A^{(0)}_n[\psi,\M_{\E}](\alpha_\E)\geq 0\, .
\end{align} 
This method also allows for the usual positivity bounds in the absence of electromagnetism to be smoothly recovered. 

Importantly, the full IR arc $\A_n[\psi,\M_{\E}]( \alpha_\E,\alpha)$ is IR-finite and calculable within the EFT, so \eqref{eq:zeroordpos} actually establishes positivity bounds on the full arc to all orders in $\alpha_\E$ (and possibly also in the Wilson coefficients), and at zeroth order in $\alpha$ since $\A_n$ and $\A_n^{(0)}$ are equivalent up to $O(\alpha)$ corrections:
\begin{equation}\label{bordal}
 \A_n[\psi,\M_{\E}](\alpha_\E,\alpha)+O(\alpha) \geq 0\,. 
\end{equation}
In an experiment (with finite $\alpha$ and a particular choice of $\E$), these $O(\alpha)$ corrections may be sizable, but we emphasize that they are, in particular, IR finite. Therefore, by improving the detector energy  resolution $\E$, we can ensure that the $O(\alpha_\E)$ bounds from \eqref{eq:zeroordpos} dominate over the $O(\alpha)$ corrections.  Finite--size experiments with $\E\neq0$ necessarily result in approximate bounds, up to the error $O(\alpha)$ or $O(G)$ that we quote. Improving on this accuracy requires going beyond the method developed in this work.    \\

The discussion of this section extends to the gravitational case as well, including the treatment of $1/t$ singularities. 
The most notable difference is the presence of strong coupling when the contour of integration in the UV representation of the arc reaches the regime $Gs\gg1$. 
 As discussed around \eqref{froisgrav}, this is not an obstruction to deriving finite dispersion relations---in particular Eq. \eqref{eq:integraldiscgeneric0} holds.

 Here, the analog of small $\alpha$ in QED is that gravity is weakly coupled  in the EFT,~$GM^2\ll1$,
\begin{equation}
\label{eq:expandingfunctionalsgravity}
\A_n[\psi,\M_{\E}](G_\E,G)= \A^{(0)}_n[\psi,\M_{\E}](G_\E)+ G M^2\, \A^{(1)}_n[\psi,\M_{\E}](G_\E)+\ldots  \,.
\end{equation}
Positivity bounds are obtained analogously to QED and take the form,
\begin{equation}
\label{eq:summaryGravityArc}
\A_n[\psi,\M_{\E}](G_\E,G)+O(GM^2) \geq 0\,. 
\end{equation}

These bounds are conceptually similar to those obtained 
in large-$N$ theories~\cite{Albert:2022oes,Fernandez:2022kzi,Berman:2023jys}, which hold only up to $O(1/N^2)$ corrections, analogous to our $O(\alpha)$ corrections.  
The situation here is perhaps more satisfactory, since increasing the ratio $\alpha_\E/\alpha$ does not require changing the theory itself, $N$,  but rather adjusting the experimental resolution $\E$ that probes the same underlying dynamics. \\

Before ending this section, it is instructive to contrast quantitatively the
obstruction to exact positivity bounds—the amount of negativity—found in
earlier works
\cite{Chang:2025cxc,Caron-Huot:2021rmr,Henriksson:2022oeu,Caron-Huot:2022ugt}
with our finding, where the negativity is finite and of order $O(GM^{2})$ or
$O(\alpha)$. 
In our language, the gravitational negativity of arcs identified in those
analyses, may be quantified by an integral (to be further integrated in $s$) analogous to
\eqref{eq:mismatchInt},
\begin{equation}\label{eqjulio}
\int_0^\E dq\, \overline{\mathrm{Im}}\,\M_{\lambda}^{\mathrm{hard}}
= F\!\left(Gs\,\log\!\frac{\E}{\lambda},\,Gs\right),
\end{equation}
where $\lambda$ is an IR regulator that is eventually sent to zero.
Positive functionals, in the sense of
\eqref{eq:positivefunctionals}, 
require that  the integral extends down to $q=0$. 
In this expression, the function $F$ is such that the limits $G\!\to\!0$ and 
$\log(\E/\lambda)\!\to\!\infty$ do not commute: one ordering
is trivial, 
while the opposite   yields a finite, calculable amount of negativity after resumming large log's   \cite{Chang:2025cxc}, but leads to a puzzling
non-decoupling effect of gravity, a point to which we return in the conclusions. 
Setting $\E\simeq \lambda$ is of no help, because while the negativity is no longer harmful, the smearing of the IR side of the dispersion reintroduce the IR sensitivity of the form $\overline{F}(G M^2\log M/\lambda)$ and the limits $G\to0$ and $\lambda\to 0$ again do not commute.

In our framework, instead, Eq.~\eqref{eq:mismatchInt} has the same form as \eqref{eqjulio}, but with the effective replacement $\lambda\to\E$, so the
integral is never logarithmically enhanced---an additional benefit of working
directly with IR--finite amplitudes. In practice, the $GM^2\log M/\E$ can be chosen to be small for a finite detector resolution $\E$.    
As a result, the correction from gravity connects smoothly to the unperturbed theory. The same
conclusion holds for electromagnetism.

\section{Positivity Bounds with Electromagnetism}
\label{Sec:BoundsEM}

In this section, we apply the general framework developed so far to the $2\to2$ scattering  of (pseudo--)Goldstone bosons arising from the spontaneous symmetry breaking pattern,
\begin{equation}
    SU(2)\times SU(2)\to SU(2)_{V} \,,
\end{equation}
with $SU(2)_V$  the diagonal subgroup. 
The resulting EFT  contains three pseudo-Goldstone bosons, $\pi^{\pm}$ and $\pi^0$. These are not exact Goldstones because
the global symmetry is weakly broken by gauging the diagonal subgroup $U(1)_V\subset SU(2)_V$ with gauge coupling $e=(4\pi\alpha)^{1/2}$, so  that $\pi_\pm$ are  charged under $U(1)_V$, while $\pi^0$ is neutral.
Moreover we consider the scenario where there is another source of explicit symmetry breaking, such that,
\begin{equation}
m_0=0\,\quad \quad m_\pm\neq 0\,,
\end{equation}
with $m_{\pm}$  finite as $\alpha\to0$ (we comment on the case $m_{\pm}\to 0$ as $\alpha\to0$ at the end of Section~\ref{sec:pipluspizero}). 
This situation is simpler to analyze because we avoid the collinear singularities associated with massless charged particles. Additionally, this scenario better captures real-world QCD.  
  Throughout this section we neglect gravity, which we discuss separately in Section~\ref{sec:positivitygravity}.

In the $SU(2)_V$ symmetric limit, the elastic scattering amplitudes % in the all--incoming notation 
are related:
\begin{equation}
\label{eq:su2symmtricamp}
\begin{split}
\M_{+0}^{SU(2)} & = M(s,t,u) \\ 
\M_{00}^{SU(2)} & =M(s,t,u)+M(t,s,u)+M(s,u,t) \\
\M_{++}^{SU(2)} & =M(s,t,u)+M(s,u,t)
\end{split}
\end{equation}
for an $s$--$u$ symmetric function, $M(s,t,u)=M(u,t,s)$.

\subsection{$\pi^+ \pi^0$ Scattering} \label{sec:pipluspizero}
We start with the process $\pi^+\pi^0\to\pi^+\pi^0$: since $\pi_0$ is neutral, the amplitude has no $1/t$ pole, making the counting in powers of $\alpha_\E$ transparent, see Section~\ref{subsubsect:functionals}.
For this process  the tree-level EFT amplitude is an $s,t,u$-analytic function symmetric in $s$--$u$ in the $\alpha\to 0$ limit and can be expanded in powers of $t$ and one independent linear combination of $su$ and $s^2+u^2$; we chose $(s-m_{\pm}^2+t/2)^2=-su/2 +(s^2+u^2)/4$ for later convenience and write,
\begin{equation}
\label{eq:suansatztree}
\M_{+0} 
^{\mathrm{tree}} =c_{0,0}+c_{1,1} t + c_{2,0}(s-m_{\pm}^2+t/2)^2+ c_{2,2}t^2+c_{3,1}t (s-m_{\pm}^2+t/2)^2 +\ldots %+ c_{4,0} (s-m_{\pm}^2+t/2)^4+\ldots 
\end{equation}
%$s_0=m_{\pm}^2-t/2$
This receives contributions  from symmetry-preserving interactions as well as from symmetry-breaking ones, such as masses and non-derivative quartic couplings.

Eq.~\eqref{eq:suansatztree} is modified by the presence of electromagnetism, and vanishes in the unregulated limit $\epsilon\to 0$. Instead, the stripped amplitude  
$
\M_{\E,+0}\equiv\lim_{\epsilon\to 0^+}{\M^{\epsilon}_{+0}}/{\W^{\mathrm{+0}} }$ is finite, with
the analytic version of Weinberg soft--exponential:
\begin{equation}\label{Wexp0piu}
    \W^{+0}(t)=\mathrm{exp}\left[\frac{\alpha }{2\pi}\frac{\left(\E^2/\mu^2\right)^{\epsilon}}{\epsilon}\left(1+2\frac{t-2m_{\pm}^2}{\sqrt{t(-t+4m_{\pm}^2)}} \arctan\frac{\sqrt{t}}{\sqrt{-t+4m_{\pm}^2}}\right)\right] \,.
\end{equation}
This depends only on $t$, confirming the arguments of Section~\ref{subsec:properties}  that the stripped amplitude shares the Regge scaling with the original amplitude.

Eq.~\eqref{Wexp0piu} is an all-order result in $\alpha_\E$. To unambiguously identify the relevant scales that enter the result,
we can work in the perturbative regime $\alpha_{\E}\ll 1$ and include the full 1-loop $O(\alpha)$ effects. 
For illustrative reasons and simplicity, we will  assume that loop effects induced by the Wilson coefficients -- thoroughly studied in Refs.~\cite{Bellazzini:2020cot,Bellazzini:2021oaj,Beadle:2024hqg,Beadle:2025cdx,Chang:2025cxc,Peng:2025klv} -- are small compared to $\alpha_\E$: $\alpha\ll g_*^2\ll\alpha_{\E}$, where $g_*^2\sim c_{i,j}M^{2i}$ denotes the generic EFT coupling strength.

The resulting stripped amplitude reads,
\begin{equation}
\label{eq:Mlambda1looppi0piplus}
\begin{split}
    \M_{\E,+0}^{(0)}(s,t)=\M_{+0}^{\mathrm{tree}}
    &\left[1+\frac{\alpha}{2\pi}\log\left(\frac{m_{\pm}^2(4m_{\pm}^2-t)}{\E^4}\right)\frac{t-2m_{\pm}^2}{\sqrt{t(4m_{\pm}^2-t)}}\arctan\left(\frac{\sqrt{t}}{\sqrt{4m_{\pm}^2-t}}\right)+\right.\\
    &\left.+\frac{\alpha}{2\pi}\log\frac{4\pi\mu^2}{\E^2}\right]+O(\alpha_\E^2)
\end{split}
\end{equation}
where the scales $m_{\pm}^2(4m_{\pm}^2-t)$ appearing in the first logarithm are exact at 1-loop, but are technically only $O(\alpha)$ and could in principle be modified by a computation of $\M_{\E,+0}$ at next-to-leading order in $\alpha$, which we do not do here.

\vspace{5mm}

To obtain positivity bounds, we follow the arguments of Section~\ref{sect:dispersion_relations} for  fixed-$t$ dispersion relations~\eqref{fixedtstuff}, with subtractions at $s_0=m_{\pm}^2-t/2$ and $C$ a contour in $s\in \mathbb{C}$ with radius $M^2-2m_\pm^2$ and center in $-t/2$, see also Appendix~\ref{appendix:num}.  In their IR representation these read,
\begin{equation}
\label{eq:arcIRpi0piplus}
\begin{split}
  \A%^{(0)}
  _n[\psi,\M_{\E,+0}]&=%\lim_{\substack{\alpha\to 0 \\ \E \to 0 \\ \alpha\log(M/\E)\to \alpha_{\E}}}
  \int_{\E}^{q_{\max}} dq\,\psi(q)\oint_{C}\frac{ds}{2\pi i}\frac{\M_{\E,+0}(s,-q^2)}{(s-m_{\pm}^2-q^2/2)^{3+n}} \,
\end{split}
\end{equation}
and satisfy
$
\A_n[\psi,\M_{\E,+0}]+O(\alpha) \geq 0$ at all orders in $\alpha_\E$, for functionals $\psi$ that respect \eqref{eq:positivefunctionals} and give a positive smeared arc in the UV representation.

We want to show how bounds on the EFT Wilson coefficients in \eqref{eq:suansatztree} change when the QED coupling is turned on. 
The bounds in the strict $\alpha=\alpha_\E=0$ limit are obtained from the tree-level arcs computable from \eqref{eq:suansatztree},
\begin{equation}\label{eq:pippi0treearc}
   \A%^{(0)}
  _n[\psi,\M_{+0}^{\mathrm{tree}}]
= \int_{\E}^{q_{\max}} dq\,\psi(q)\left( c_{2,0}-q^2 c_{3,1}+\ldots\right)\,.
\end{equation}
giving\footnote{An upper bound on  $c_{3,1}/c_{20}$ can also be derived, but only in terms of the couplings of other elastic channels, $\pi^0 \pi^0$ and $\pi^+ \pi^+$. In the $SU(2)_V$ limit, one can relate these couplings to the Wilson coefficients of $\pi^+ \pi^0$.}
\begin{equation}
\label{eq:boundspi+pi0tree}
c_{2,0} >   0  \,,\qquad  
(M^2-m_{\pm}^2)\frac{c_{3,1}}{c_{2,0}}\geq -\frac{3}{2} \,.
\end{equation}
The smearing functions $\psi$ associated to these bounds satisfy simultaneously \eqref{eq:positivefunctionals}, as well as orthogonality to higher Wilson coefficients, via $\int_{\E} dq \psi(q) q^{2n}=0$ for $n\geq 2$, in order for the bound to be independent from them.
For $n\to\infty$ such functionals
% have support only at $q=\E$~\cite{Beadle:2024hqg} and 
tend to linear combinations of $\delta(q)$ and $\delta^\prime (q)$ as $\E\to 0$ \cite{Beadle:2024hqg}, and the problem becomes equivalent  to matching $q$-derivatives in  UV and IR.

\subsubsection*{Conservative bounds}
Now we work at finite $\alpha_\E$, where bounds obtained with ordinary amplitudes are moot.
With stripped amplitudes, the $\pi^+\pi^0$ case is somewhat special since the $\alpha,\E\to0$ limits commute with the $q$-integral, thanks to the absence of  non-integrable $1/q^2$ contributions.
In fact, the loop factor in Eq.~\eqref{eq:Mlambda1looppi0piplus} is analytic at $t=0$ and arcs  can once again be expanded in powers of $q^2$ in terms of effective coefficients $C_{i,j}$,
\begin{equation}
    %\oint_{C}\frac{ds}{2\pi i}\frac{\M_{\E,+0}^{1\text{-loop}}(s,-q^2)}{(s-m_{\pm}^2-q^2/2)^{3}}
    \A^{(0)}
  _n[\psi, \M_{\E,+0}]
=  \A%^{(0)}
  _n[\psi, \M_{\E,+0}^{(0)}]
= \int_{\E}^{q_{\max}} dq\,\psi(q)\left(C_{2,0}-q^2 C_{3,1}+\ldots\right)\,.
\end{equation}
where the $C_{k+2,k}$ depend on  $c_{i+2,i}$ with $i\leq k$, as well as on $\alpha$ and $m_{\pm}$: they can be explicitly computed from Eqs. \eqref{eq:suansatztree} and \eqref{eq:Mlambda1looppi0piplus}.

At this point, without any further assumption on the EFT coefficients (besides weak coupling),  the same smearing functions that were used at tree-level can be used to derive positivity bounds which include the leading electromagnetic corrections, by simply replacing $c_{i+2,i}\to C_{i+2,i}$ in \eqref{eq:boundspi+pi0tree}.
At first order in $\alpha_\E$, and in the limit $m_{\pm}\ll M$, %(exact expressions appear in Appendix~\ref{} \fr{do they?})
\begin{equation}
\label{eq:boundspi+pi0explicit}
\begin{split}
c_{2,0} +O(\alpha_\E^2,\alpha) & \geq   0  \\ 
c_{2,0}\left[\frac{3}{2}-\frac{(M^2-m_{\pm}^2)}{m_{\pm}^2}\frac{\alpha}{3\pi} \log(\frac{\E}{m_{\pm}})\right]+ (M^2-m_{\pm}^2) c_{3,1}  +  O(\alpha_\E^2,\alpha) &\geq 0 \,.
\end{split}
\end{equation}
 The exact 
dependence of bounds on $m_{\pm}$ is illustrated in Fig.~\ref{pi0pipmpivar}.
Crucially, all terms are IR-finite, 
with the electromagnetic corrections weakening the tree--level bound on $c_{3,1}$,
and converging to the familiar \eqref{eq:boundspi+pi0tree}
as $\alpha,\alpha_{\E}\to 0$.

These results illustrate that the presence of long-range electromagnetic interactions does not spoil the usual analyticity--based positivity properties of the low-energy EFT. Rather, once the appropriate IR-finite stripped amplitudes are used, the bounds retain the same structure as in short-range theories, up to calculable logarithmic corrections controlled by $\alpha_{\E}$. 

The constraints  \eqref{eq:boundspi+pi0explicit} are continuously connected to the constraints that hold with no electromagnetic interactions as long as $m^2_{\pm}$ remains finite in the limit $\alpha\to0$. 
It is interesting to consider also the scenario where $m_\pm$ is not fixed, but $m_{\pm}^2\propto M^2 \alpha/4\pi $, corresponding to a situation where electromagnetic interactions are the only source of explicit symmetry breaking.
In this scenario, the bounds appear to change discretely. This is  not very surprising since the IR mass singularities are correlated with the coupling constant\footnote{A similar phenomenon indeed happens in the EFT of spin--3/2 particles of mass $m_{3/2}$, which is found inconsistent with positivity bounds whenever $M\gg m_{3/2}$, unless gravitons are also present and couple in a supersymmetric way \cite{Bellazzini:2025shd} even in the decoupling limit $G\to0$, since  $m_{3/2}^2\propto G$.} $1/m_{\pm}^2\sim 1/\alpha$;
however, in this massless limit, collinear singularities -- as manifest by the branch point of $\W^{+0}(t)$ at $4m_{\pm}^2\to0$ -- would become important and our calculation would have to be rethought. Nevertheless, as long as one restricts to a detector resolution $\E$ as small as $\sqrt{\alpha} M \mathrm{exp}(-\alpha_{\E}/\alpha)$ with $\alpha_{\E} \ll 1$, it's still possible to perform a perturbative calculation  within the framework developed here. It then follows again that $c_{3,1}M^2 /c_{2,0}\geq -O(1)$.

\subsubsection*{EFT-informed bounds}\label{eft-informed}
 
\begin{figure}[t]
\begin{centering}
\includegraphics[width=0.6\textwidth]{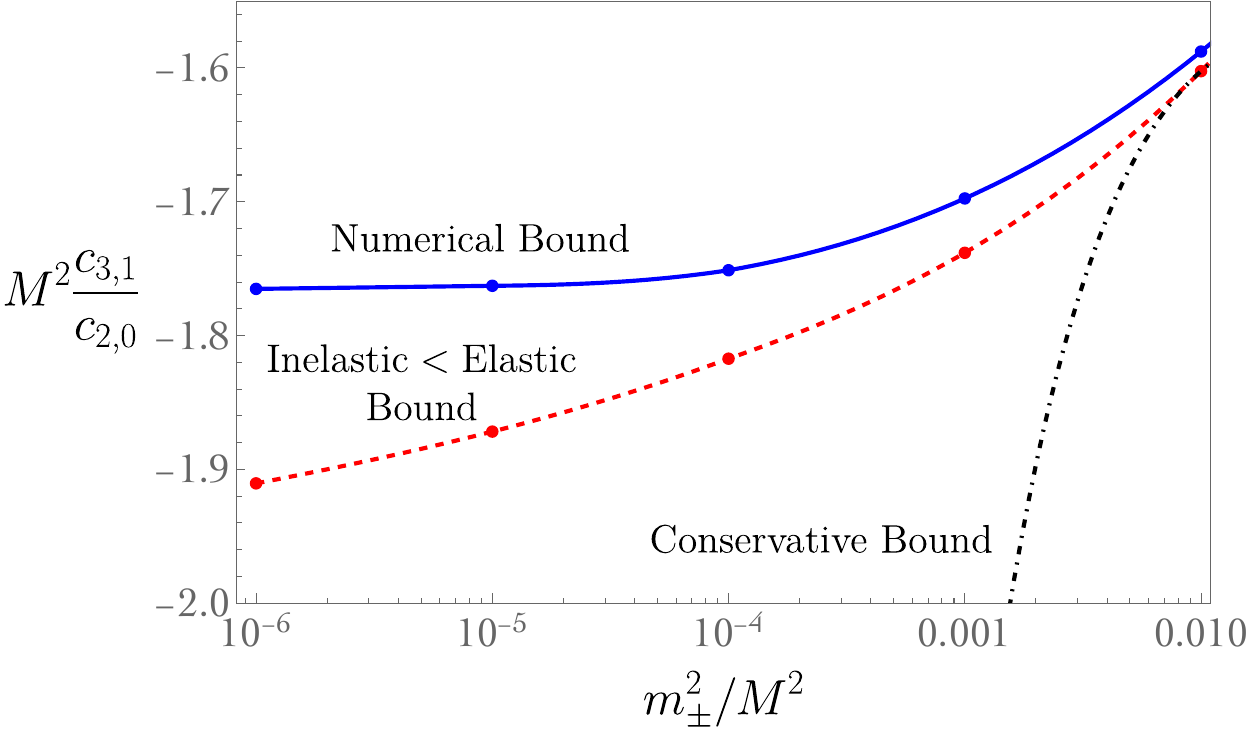}
  \caption{Lower bounds on $M^2c_{3,1}/c_{2,0}$ as function of $m_\pm^2/M^2$, for fixed  $\alpha_{\E} =1/100$. Conservative bounds from \eqref{eq:boundspi+pi0explicit}   (black, dot-dashed) have no assumptions on higher coefficients, semianalytic bounds \eqref{eq:semianalytic}  (red dashed) ignore EFT terms apart from $c_{2,0}$ and $c_{3,1}$.
  The numerical bounds (blue) ignore only terms larger than $N$, controlling its convergence up to  $N=14$. }
    \label{pi0pipmpivar}
    \end{centering}
\end{figure}
The previous analysis relied on a generic parametrization of the low-energy amplitude $\M_{+0-0}^{\mathrm{tree}}$, in which Wilson coefficients could take any value prior to the positivity bounds -- this is how tree-level bounds are traditionally derived.
This (unrealistic) setup becomes untreatable beyond tree-level, where loops of very irrelevant couplings  contribute also to the lower arcs, and one has to \emph{assume} they be smaller than the tree level coefficients  $ |c_{n,k}|\, M^{2n} / 16 \pi^2\lesssim 1$  for the ansatz to be consistently calculable  \cite{Bellazzini:2020cot}.
In our specific case, the problem of hypotheses-less  generic amplitudes is that they include also those in which the  coefficients $c_{n,k}$ (controlled by UV physics at the scale $M$) completely \emph{undo} the radiative IR effects of electromagnetism, associated with the scale~$m_\pm$. 

A very mild assumption about the asymptotic growth of Wilson coefficients will circumnavigate this issue, and reflects the familiar situation in which marginal interactions, albeit weak, always  dominate  over sufficiently irrelevant operators.
Scale separation, $m_{\pm}\ll M$, and the requirement that 
\begin{equation}\label{eq:ass}
 % |c_{n,k}|\, M^{2n} \lesssim 1\,,
|c_{n,k}|\ll \frac{4\pi\alpha_\E}{m_{\pm}^{2n}}\,,\quad \textrm{for}\quad n\geq N
\end{equation}
for a sufficiently large $N$, does the job. Notice that for sufficiently large $N$ such an assumption is even weaker than requiring perturbative control within the EFT; moreover it is always   satisfied if the bounds in the exact $SU(2)$ limit also hold.
We formalize this assumption by truncating the EFT ansatz of Eq.~\eqref{eq:suansatztree} at some     order $N$ and neglect $c_{n,k}$ for $n\geq N$, while retaining the exact functional dependence induced by IR QED effects in Eq.~\eqref{eq:Mlambda1looppi0piplus}. 

If we neglect all terms larger than $N=4$, bounds of this type can be obtained semianalytically by directly studying the $q$-integrand of arcs as a function of $q$.  The UV-representation of arcs implies that the inelastic $q\neq 0$ part is  bounded by the elastic  $q=0$ one, at the level of the  $q$-integrand -- this follows simply because Legendre polynomials are bounded by~$|P_\ell|\leq 1$.  In the IR part, ignoring EFT terms with $n\geq N=4$ we have only $c_{2,0}$ and $c_{3,1}$, and we obtain a simple inequality~\cite{Bellazzini:2023nqj,Bellazzini:2025shd},
\begin{equation}
\label{eq:semianalytic}
\left|\frac{L(q)}{L(0)}(1+t\,c_{3,1}/c_{2,0})\right|\lesssim \left(\frac{M^2-m_{\pm}^2}{M^2-m_{\pm}^2-q^2/2}\right)^3\,,\qquad  q<q_{\mathrm{max}}\,, 
\end{equation}   
with $L(q)=\M_{\E,+0}/\M^{\mathrm{tree}}_{\E,+0}$ collecting the soft factor and leading-$\alpha$ effects appearing in the square bracket in \eqref{eq:Mlambda1looppi0piplus}, and the right-hand side capturing the $q$-modulation of the UV integrand maximizing $|P_\ell|\to 1$. 
 We show the result, as a function of $m_\pm^2/M^2$ for fixed  $\alpha_{\E}$  
 as a red line in Fig.~\ref{pi0pipmpivar}. Notice how the additional information provided by \eqref{eq:ass} leads, at small pion mass, to bounds that are more restrictive than the agnostic ones from \eqref{eq:boundspi+pi0explicit}.

\begin{figure}[t]
\begin{centering}
    \includegraphics[width=0.6\textwidth]{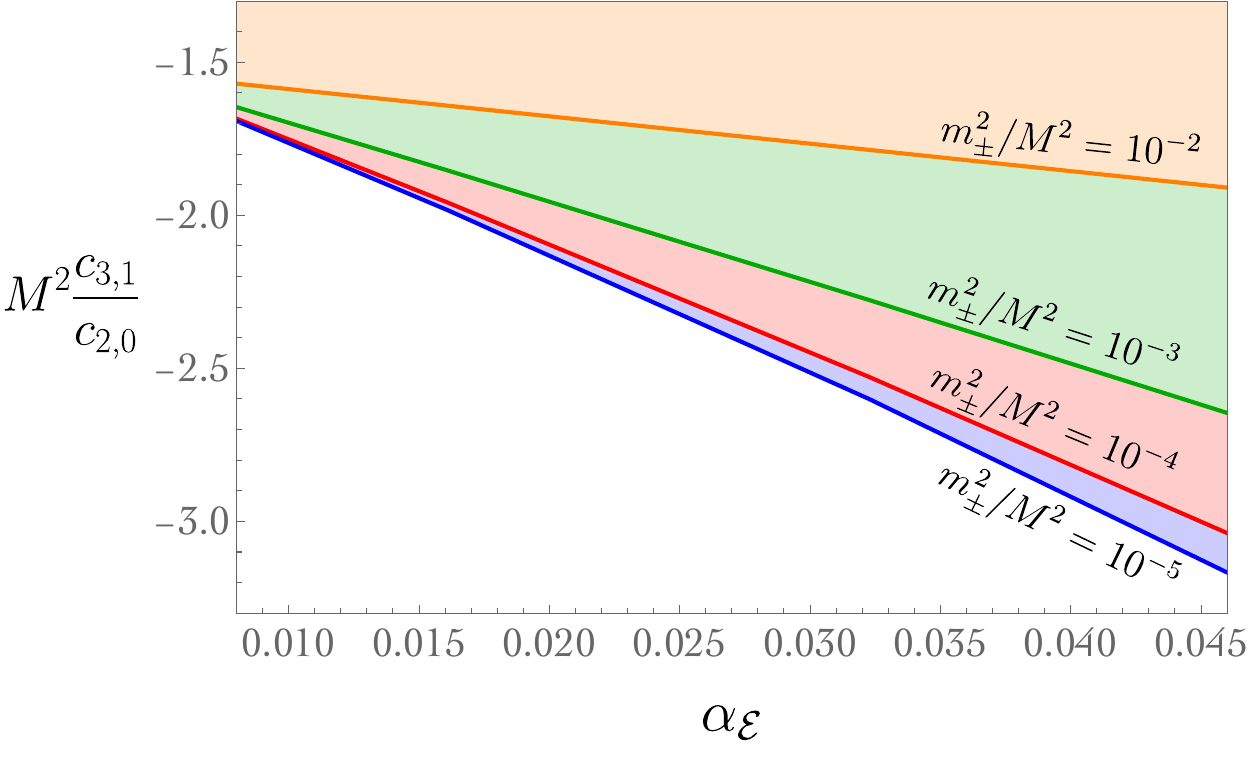}
     \caption{Numerical---EFT--informed--- lower bound  of $M^2c_{3,1}/c_{2,0}$ as a function of $\alpha_ \E = \alpha \log(M/\E)$ for various  values of $m_{\pm}^2/M^2$.   As $\alpha_\E$ is lowered, the numerical bounds approaches the tree level one, given in \eqref{eq:boundspi+pi0tree}. A fit for $m^2_{\pm}/M^2=10^{-2}$ is given in \eqref{eq:fitpi0pi+}.    }
    \label{pi0pipalphavar}
    \end{centering}
\end{figure}

For larger $N$, we study the problem numerically as explained in Appendix~\ref{appendix:num}.
To remove sensitivity to the coefficients $c_{n,k}$ with $3<n< N$, we use functionals satisfying positivity,  \eqref{eq:positivefunctionals},  as well as  the condition $
\int_{\E}^{q_{max}}dq\psi(q) q^{2n} L(q)=0$.
 We then study the asymptotic behavior of positivity bounds  as $N$ is increased up to $N=14$. The results appear in blue in Fig.~\ref{pi0pipmpivar}, providing a refinement of the semi-analytic bound of \eqref{eq:semianalytic}.

In Fig.~\ref{pi0pipalphavar}
we study numerically the behavior of the bounds as one varies $\alpha_\E$ for different fixed values of $m_\pm^2/M^2$. As expected, in the limit $\alpha_\E \to 0$ we recover the tree level bound in \eqref{eq:boundspi+pi0tree}, up to  subleading $O(\alpha)$ effects. Taking $m_\pm^2=10^{-2}M^2$ as example, a fit of the numerical bounds gives:
\begin{equation}
\label{eq:fitpi0pi+}
M^2c_{3,1}/c_{2,0} \geq  -1.498-8.937 \alpha \log(M/\E)\qquad (m^2_{\pm}/M^2=10^{-2}) 
\end{equation}

\begin{figure}[t]
\begin{centering}
\includegraphics[width=0.6\textwidth]{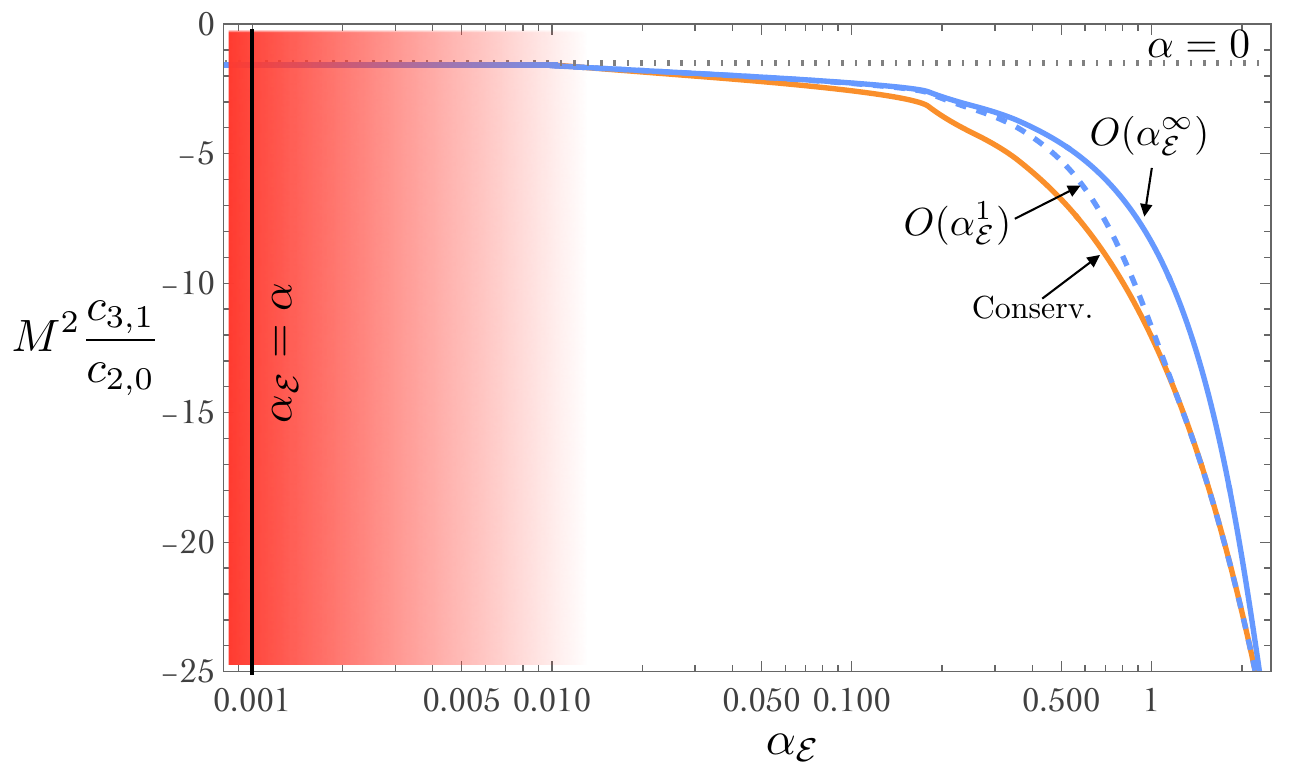}
     \caption{Lower bound on $c_{3,1}/c_{2,0}$ as a function of $\alpha_\E$ obtained from the all-order-$\alpha_\E$ result (blue, solid) and the first order result (blue, dashed), compared with the bound in absence of QED (dotted) and the conservative bound of \eqref{eq:boundspi+pi0explicit} (orange); $m_\pm=0.1M$ and $\alpha=10^{-3}$. For small $\alpha_\E\sim \alpha$, our computational framework breaks down, as illustrated by the red shading. For too large $\alpha_\E$, instead, the  bound deteriorates. The sweet spot lies at moderate   $\alpha\ll \alpha_\E\lesssim 1$, where the linear approximation holds. }
    \label{quali}
    \end{centering}
\end{figure}

Our derivation allow us  to have control of the results in the opposite regime, $\alpha_{\E}\gtrsim 1$, by working with the full Weinberg exponential. 
 Fig.~\ref{quali} shows the $\alpha_\E$-dependence of the lower bound, derived with the all-order and the linear in $\alpha_\E$ results. At large $\alpha_\E$, the conservative and EFT-informed bounds asymptote to each other.
 
 This figure also shows the tension between taking small $\alpha_\E\to \alpha$  (where our computation fails) and taking larger $\alpha_\E$, where the bounds become less and less stringent.
 The physical interpretation of $\E$ allows us to chose detectors with of right size
 to apporach the best bounds at  intermediate values of $\alpha_\E$.

\subsection{$\pi^+ \pi^+$ Scattering}\label{sec:pipluspiplus}

We now consider the scattering process $\pi^+\pi^+ \to \pi^+\pi^+$
within the same EFT.
As in the $\pi^+\pi^0$ case, we work with stripped amplitudes in order to
construct a well-defined analytic function free of IR divergences.
There is, however, one important qualitative difference with respect to
$\pi^+\pi^0$ scattering: the presence of a $t\to0$ singularity,
which would naively obstruct the derivation of positivity bounds in
$D=4$~\cite{Caron-Huot:2021rmr,Henriksson:2022oeu,Caron-Huot:2022ugt}.\footnote{The $1/t$  tree level pole does not enter twice-subtracted
dispersion relations. At loop level, however, box-type diagrams involving
only electromagnetic interactions generate both a singular behavior in $q$  and
non-analyticity in $s$ thereby propagating the singularity to all
$n$-subtracted dispersion relations in the schematic form of  $(\log q)/q^2$ terms.}
As explained in Section~\ref{sect:dispersion_relations}, working with
IR--finite stripped amplitudes automatically resolves this issue: the
$1/t$ singularity gives rise only to a finite, calculable amount of
negativity, as we explicitly show here for  $\pi^+\pi^+$-scattering.

The amplitude is $t$--$u$ symmetric and can be expressed in terms of the  variable $s$ and $q^2=tu/s$.   
At tree level, it can be written as,
\begin{equation}\label{trepiomeno}
\M_{++}^{\mathrm{tree}} = 8\pi\alpha \left(1-\frac{s^2-6s m_{\pm}^2+8m_{\pm}^2}{q^2s}\right) + \bar{c}_{0,0}+s\bar{c}_{1,0}+s^2\bar{c}_{2,0}+\bar{c}_{2,1}sq^2+\bar{c}_{3,0}s^2+\bar{c}_{3,1}s^2q^2+\ldots
\end{equation}
where, in the $SU(2)_V$ (and massless) limit, $\bar{c}_{i,j}$  are linear combinations of $c_{i,j}$  from $\pi^+\pi^0$ scattering in \eqref{eq:suansatztree}.

Beyond tree-level, the amplitude is plagued by the IR effects captured by~\eqref{eq:AnalyticSoftQED}, 
\begin{equation}
\begin{split}
 \!\!
    \W^{++}(s,q)=\mathrm{exp}\Bigg\{\frac{\alpha }{\pi}\frac{\left(\E^2/\mu^2\right)^{\epsilon}}{\epsilon}\Bigg[1-\Bigg(2&\frac{s-2m_{\pm}^2}{\sqrt{s(-s+4m_{\pm}^2)}} \arctan\frac{\sqrt{s}}{\sqrt{-s+4m_{\pm}^2}}\\
    &~~~~~~~~-(s\rightarrow t_{+}(s,q))-(s\rightarrow t_-(s,q))\Bigg)\Bigg]\Bigg\} \,,
\end{split}
\end{equation}
where $t_{\pm}$ correspond to $t$ and $u$ in terms of $s$ and $q$, see Appendix \ref{appendix:TUSymm}.
Analogously to the amplitude itself,
this exponential  satisfies  $s$--$u$ crossing  $\W^{++}(u(s,q),t(s,q))=\W^{+-}(s,t(s,q))$ into the  process $\pi^+\pi^-\to\pi^+\pi^-$.

The IR-finite stripped amplitude
$\M_{\E,+\pm}=\lim_{\epsilon\to 0^+}{\M^{\epsilon}_{+\pm}}/{\W^{\mathrm{+\pm}}}$ has the same 
Regge scaling as the original amplitude, as can be seen by expanding the exponentials at $s\gg q^2$.
We compute the stripped amplitude explicitly at one-loop and report its full version in Appendix~\ref{appendix:pi+pi+}.

Positivity bounds in this context are conveniently derived at real fixed $q^2=tu/s$, via $t$-$u$ crossing symmetric dispersion relations~\cite{Bellazzini:2025shd}, which follow the structure of Section~\ref{sect:dispersion_relations} but are also explicitly expanded upon in Appendix~\ref{appendix:TUSymm}.
Their appeal lies in the fact that, at tree-level,  only finitely many Wilson coefficients enter into any IR arc. For instance,
\begin{align}\label{eq:plusplusA0}
    \A%^{(0)}
  _0[\psi, \M^{\textrm{tree}}] = \int dq\,\psi(q)(c_{2,0} + c_{3,1}q^2+ c_{4,2}q^4)\,,
\end{align}
which has to be contrasted with the $\pi^+\pi^0$ case with fixed-$t$ dispersion relations, in which infinitely many coefficients appear in every arc, see \eqref{eq:pippi0treearc}.

Matching to the UV representation gives simple tree--level sum rules for the Wilson coefficients~\eqref{eq:treelevelpi+pi+Sumrules}.
From smearing functions $\psi$ that tend (for $\E\to 0$) to linear combinations of $\delta(q)$ and~$\delta'(q)$  the simplest bounds on the leading coefficients are,
\begin{equation}
\label{eq:treepi+pi+bounds}
\textrm{Tree-level:}\quad\bar{c}_{2,0}\geq 0\,, 
\qquad  
-\bar{c}_{2,0}/(M_u^2-4m_{\pm}^2) \leq \bar{c}_{3,0} \leq \bar{c}_{2,0}/M_{s}^2\,, 
\qquad  
M_u^2\,\bar{c}_{3,1}< 3\bar{c}_{2,0}\,,
\end{equation}
where $M_s^2$ and $M_u^2$ denote the lowest thresholds for new states in $\M_{++}$ and $\M_{+-}$.  
\vspace{3mm}

To incorporate radiative effects, we follow the strategy of the previous subsection.   
Differently from $\pi^+\pi^0$, here smearing does not commute with the scaling limit \eqref{eq:generalscaling}, because of the $q\to 0$ singular behavior of the amplitude from QED -- see Section~\ref{subsubsect:functionals}. For instance, the $n=0$ arc (in the limit $M_s=M_u\equiv M \gg q_{\max},m_{\pm}$),
including the tree--level contributions together with the dominant photon loop, reads,
\begin{equation}
\label{eq:A0approxlimit}
 \A^{(0)}
  _n[\psi, \M_{\E,++}]=\lim_{\substack{\alpha\to 0 \\ \E \to 0 \\ \alpha\log(M/\E)\to \alpha_{\E}}}
\int_{\E}^{q_{\max}} dq\,\psi(q)
\left[
\bar{c}_{2,0} + \bar{c}_{3,1} q^2 + \bar{c}_{4,2} q^4
+ \frac{8 \alpha^2 }{M^2 q^2}\log \!\left(\frac{q^2}{4\pi\,\E^2}\right)
\right]\,.
\end{equation}
For simplicity, and just to highlight the impact of photon loops, we restrict our
analysis to the clean configuration of couplings satisfying
 $ \bar{c}_{n,k}M^{2n} \ll \alpha_{\E}$, which is in particular compatible with $\alpha_{\E}^2\sim |\bar{c}_{n,k}|^{2n}$.

\begin{figure}[h!!]
\begin{centering}
    \includegraphics[width=0.49\textwidth]{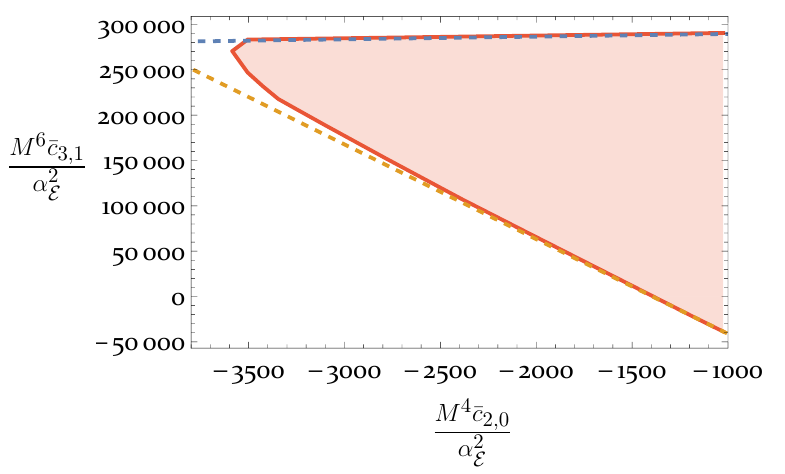}
    \includegraphics[width=0.49\textwidth]{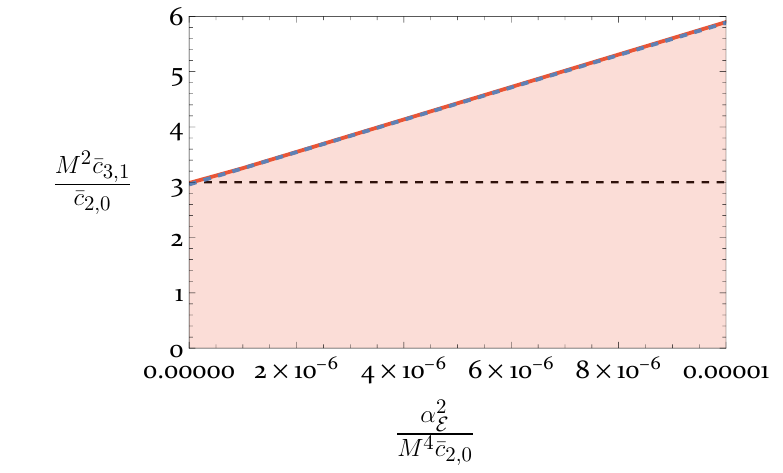}
     \caption{Left panel: one-loop bounds on $M^6\bar{c}_{3,1}/\alpha_\E^2$ and $M^4\bar{c}_{2,0}/\alpha_\E^2$.  In red the one--loop allowed region. The dashed lines are obtained by optimizing only on the $\bar{c}_{2,0}$ term and fixing $\psi'(0)$ to $18280$ (blue), $4\cdot10^4$ (orange), while keeping fixed $\int_{\E}^{q_{max}}dq\,\psi(q)q^2=\pm1$, and respectively correspond to the upper and lower bound in Eq. \eqref{eq:boundsc31pi+pi+}. Right panel: one-loop upper bound on $M^2\bar{c}_{3,1}/\bar{c}_{2,0}$ as a function of $\alpha_{\E}^2/(M^4\bar{c}_{2,0})$. The dashed blue line corresponds to the upper bound in Eq. \eqref{eq:boundsc31pi+pi+}, while the black dashed line is the tree-level bound.}
    \label{fig:boundspipluspiplus}
    \end{centering}
\end{figure}

By selecting admissible functionals $\psi(q)$ which satisfy \eqref{eq:positivefunctionals}, we obtain positivity bounds.
In particular, by further identifying functionals that integrate to zero on the $c_{3,1}q^2$ and $c_{4,2}q^4$ terms
we  obtain,\footnote{The bound is saturated by EFT couplings $g_*^2\sim \alpha_\E^2$, compatible with our computations at  leading order in~$g_*^2$. }
\begin{align}
\label{eq:boundc20pi+pi+}
    \bar{c}_{2,0} M^4+(59.9 \alpha_{\E})^2 +O(\alpha) > 0\, .
\end{align}
By instead choosing functionals that project out only the $\bar{c}_{4,2}$ term, we get bounds on $\bar c_{3,1}$ in terms of $\bar c_{2,0}$. In Fig. \ref{fig:boundspipluspiplus} we show in red the bounds obtained through the standard numerical algorithm \cite{Caron-Huot:2021rmr}. In addition, we can obtain simple bounds with an analytic dependence on $\alpha_{\E}$ by optimizing on the $\bar c_{2,0}$ term while keeping fixed the $\bar c_{3,1}$ term and scanning different values of the normalization $\psi'(0)$. For instance,
\begin{align}\label{eq:boundsc31pi+pi+}
    \bar{c}_{3,1}M^6&\leq 3 \bar c_{2,0}M^4 + (541\alpha_{\E})^2 + O(\alpha)\,,\\[0.2cm]
    \bar c_{3,1}M^6&\geq-104\bar c_{2,0}M^4-(382\alpha_{\E})^2+ O(\alpha)\,,
\end{align}
respectively obtained with $\psi'(0)=18280$ and $\psi'(0)=4\cdot10^4$.
These linear constraints draw lines in the $\bar c_{3,1}$--$\bar c_{2,0}$ plane tangent to the allowed region, as shown in Fig. \ref{fig:boundspipluspiplus}. The bounds are manifestly continuous as electromagnetism is decoupled $\alpha_\E\to0$, recovering the tree-level results, see Fig. \ref{fig:boundspipluspiplus}.

Bounds on higher Wilson coefficients are affected in a similar way. We have not attempted to reach the most optimal constraints, partly because the regime $q_{\max}\ll M$ -- that leads to simpler integrable expressions --  restricts the support of~$\psi$, partly because our goal here is to demonstrate explicitly that IR--finite bounds exist and connect smoothly to the tree--level limits.

\section{Positivity Bounds with Gravity}
\label{sec:positivitygravity}

We discuss now the scattering of identical massless shift--symmetric scalars exclusively coupled to gravity.
The tree--level amplitude in the low energy EFT can be expanded as 
\begin{equation}
\M^{\mathrm{tree,GR}}=8\pi G\left(\frac{t u}{s}+\frac{u s}{t}+\frac{s t}{u} \right)+\hat{c}_{2,0}\left(\!\frac{s^2+t^2+u^2}{2}\!\right)+\hat{c}_{3,1}stu+\hat{c}_{4,0}\left(\!\frac{s^2+t^2+u^2}{2}\!\right)^2\!+\ldots
\end{equation}

In the gravitational case, the relevant soft factor is given by Eq.~\eqref{eq:WeinbergEqualvanishingMasses},
\begin{equation}
 \W_{\mathrm{GR}}(s,t)=\mathrm{exp}\left[-G\frac{\left(\E/\mu\right)^{2\epsilon}}{\pi\epsilon} \left(s\log\left(-s/\mu^2\right)+t\log\left(-t/\mu^2\right)+u\log\left(-u/\mu^2\right) \right) \right]\,,
\end{equation}
and, as usual, the IR-finite stripped amplitude is $
  \M_{\E}^{\mathrm{GR}}\equiv\lim_{\epsilon\to 0^+}{\M^{\epsilon}/}{\W_{\mathrm{GR}} }$.
The full one--loop stripped amplitude is reported in Appendix \ref{app:amps}.

The amplitude is $s-t-u$  symmetric:  fully crossing symmetric dispersion relations at fixed $q^2=2stu/(s^2+t^2+u^2)$ -- discussed in Appendix~\ref{app:stu} -- are the most convenient ones to use. Indeed, at tree-level, the IR representation of the $n=0$ arc is simply,
\begin{equation}
\label{eq:deltaA0QEDonly}
\begin{split}
    \mathcal{A}_0^{(0)}[\psi,\M^\textrm{tree,\,GR}]
    &= \lim_{\substack{G\to 0 \\ \E \to 0 \\ GM^2\log M/\E\to G_{\E}}}
    \int_{\E}^{q_{max}} dq\,\psi(q)\,
\left[  \frac{8 \pi  G}{q^2}+\hat c_{2,0}+q^2 \hat c_{3,1} \right] \,.
\end{split}
\end{equation}
For ordinary IR--divergent amplitudes, a similar expression is found in the literature; in that case, even at tree-level, in $D=4$ dimensions, functionals that are positive in the UV representation are not integrable against the pole in $q$ as the IR--cutoff is removed~\cite{Caron-Huot:2021rmr}. Instead, for IR--finite stripped amplitudes,  the amount of negativity in the integral in \eqref{eq:deltaA0QEDonly} is both finite and of order $O(GM^2,\E^2/M^2)$, see Section~\ref{sect:dispersion_relations}, and can therefore be discarded in the scaling limit \eqref{eq:generalscaling}.

At loop level,  the $G_\E^2$ term contributing to the arcs is obtained from the first line of the full $G^2$ amplitude in Eq. \eqref{eqspleuf}. Following \cite{Beadle:2025cdx,Chang:2025cxc}, the arc with $n=0$ evaluates to
\begin{equation}
\label{eq:deltaA0GRonly}
\begin{split}
    \delta\mathcal{A}_0^{(0)}[\psi,\M_{\E}^{\mathrm{1-loop,\, GR}}]
    &= \lim_{\substack{G\to 0 \\ \E \to 0 \\ GM^2\log M/\E\to G_{\E}}}
    \int_{\E}^{q_{max}} dq\,\psi(q)\,
   \frac{16G^2M^2}{q^2} \log (\frac{\E^2}{M^2} )  \,.
\end{split}
\end{equation}
As in the $\pi_{+}\pi_{+}$ case, just to highlight the impact of graviton loops, we restrict our
analysis to the configuration of couplings satisfying $\hat{c}_{n,k}M^{2n} \ll G_{\E}$.

\begin{figure}[t]
\begin{centering}
    \includegraphics[width=0.49\textwidth]{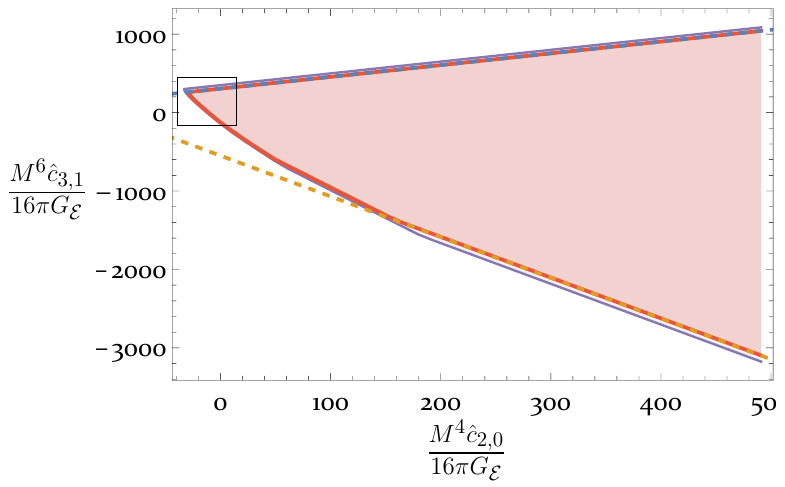}
    \includegraphics[width=0.49\textwidth]{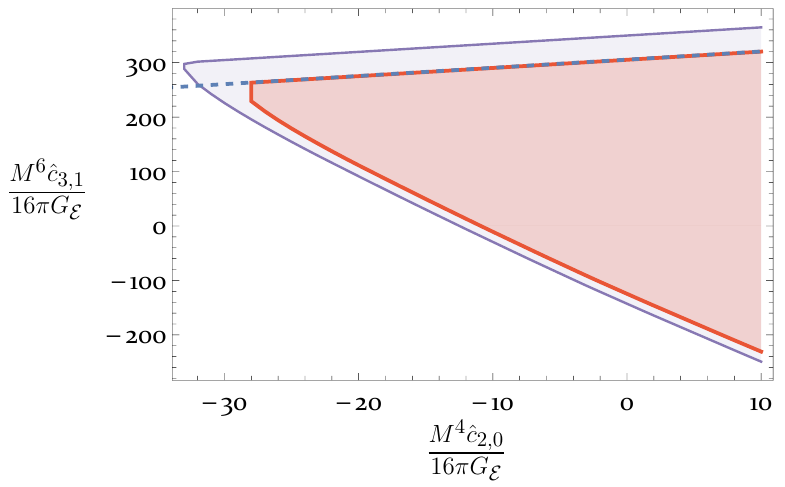}
     \caption{Bounds on $M^6\hat{c}_{3,1}/(16\pi G_{\E})$ and $M^4\hat{c}_{2,0}/(16\pi G_{\E})$. In red the one--loop allowed region ($G_{\E}=0.1$), in purple the tree--level one. The dashed lines respectively correspond to the upper and lower bound in Eq. \eqref{eq:boundsgr2}. In the right panel we zoom in on the tip of the allowed region.}
    \label{fig:boundsgr}
    \end{centering}
\end{figure}

\begin{figure}[h!!]
\begin{centering}
    \includegraphics[width=0.7\textwidth]{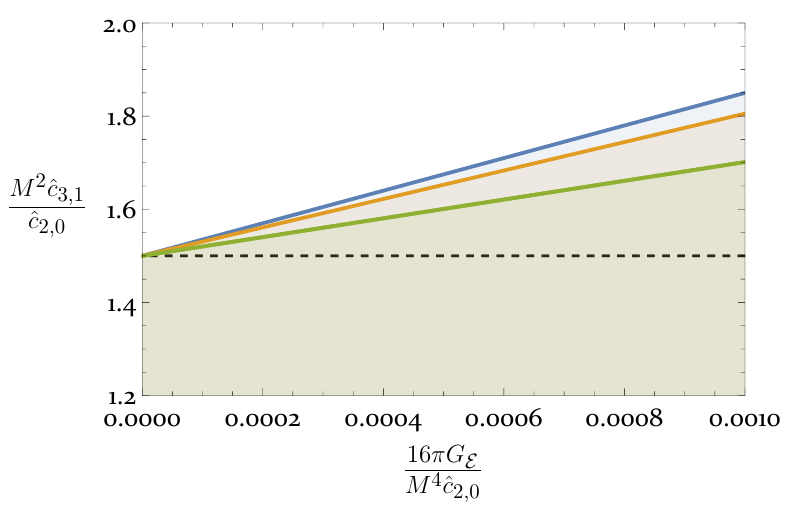}
     \caption{Upper bounds on $M^2\hat{c}_{3,1}/\hat{c}_{2,0}$ as a function of $(16\pi G_{\E})/M^4\hat{c}_{2,0}$. The blue line corresponds to the upper bound in Eq. \eqref{eq:boundsgr2} evaluated at tree--level. The orange and green lines correpond to the same bound computed at $G_{\mathcal{E}}=1/10,1/3$. The black dashed line is the tree--level bound in absence of gravity.}
    \label{fig:boundsgr2}
    \end{centering}
\end{figure}

Positivity bounds are obtained by looking for admissible functions $\psi(q)$. By requiring that the latter integrate to zero against $q^2$, removing the $\hat{c}_{3,1}$ contribution, we get
\begin{equation}\label{eq:boundsgr1}
    \hat{c}_{2,0}M^4>-65.7\left[8\pi G_{\E}-\frac{(8\pi G_{\E})^2}{2\pi^2}\right] +O(G)\,,
\end{equation}
while releasing this constraint we obtain two--sided bounds on $\hat{c}_{3,1}q^2$, following the standard procedure \cite{Caron-Huot:2021rmr,Caron-Huot:2022ugt}. Analogously to the $\pi_{+}\pi_{+}$ case, we can keep the $G_{\E}$ dependence analytic and derive constraints of the form
\begin{align}\label{eq:boundsgr2}
    \hat{c}_{3,1}M^6&\leq\frac{3}{2}\hat{c}_{2,0}M^4+700\left[8\pi G_{\E}-\frac{(8\pi G_{\E})^2}{2\pi^2}\right]+O(G)\,,\\
    \hat{c}_{3,1}M^6&\geq-5.2\hat{c}_{2,0}M^4-1260\left[8\pi G_{\E}-\frac{(8\pi G_{\E})^2}{2\pi^2}\right]+O(G)\,.
\end{align}
respectively obtained with $\psi'(0)=350$ and $\psi'(0)=630$. The bounds are shown in Fig. \ref{fig:boundsgr}. As in the $\pi_+\pi_+$ case, we did not focus on optimizing the constraints. Importantly, as $G_{\E}$ goes to zero we recover the tree-level bounds in absence of gravity $\hat c_{2,0}>0$, $\hat{c}_{3,1}M^2<\frac{3}{2}\hat{c}_{2,0}$, as highlighted in Fig. \ref{fig:boundsgr2}. Also the lower bound on $\hat{c}_{3,1}$ is consistent with the analytic tree--level bound \cite{Caron-Huot:2021rmr}.

\section{Summary and Conclusions}
\label{sect:conclusions}

In this work we developed a general framework for carving out the space of low--energy EFTs, based on unitarity and causality constraints, in theories with long--range interactions, focusing on QED and gravity in $D=4$ flat spacetime. The central obstacle in these theories is the presence of universal IR divergences, which   trivialize all scattering amplitudes and conventional dispersion--relation arguments.

To overcome this obstruction, we introduced \emph{stripped amplitudes} $\M_{\E}$, a one–parameter family of functions obtained by factoring out from standard amplitudes the analytic continuation of the Weinberg soft exponential. For any $\E >0$, the resulting $\M_{\E}$ are (exactly) IR finite, analytic, crossing symmetric, and Lorentz invariant. For~$\E$ exponentially small relative to any other hard scale $M$, this parameter acquires a clear physical meaning: it coincides with the energy resolution $\Delta E=\E$ of a macroscopic device capable of detecting soft photons or gravitons~\eqref{eq:inclusiveXsection}. Different values of $\E$ correspond to different  experimental apparatuses.  

In the scaling limit $\alpha\to 0$ or $G\to 0$ with $\alpha_{\E}=\alpha\log(M/\E)$ or $G_{\E}=GM^2\log(M/\E)$ held fixed---capturing the universal behavior of all such large detectors---the stripped amplitudes become unitary as well. In this regime, $\M_{\E}$ may be viewed as a genuine ``detector amplitude'', possessing exactly the analyticity and unitarity properties required to formulate dispersion relations and derive positivity and bootstrap bounds.~\footnote{One could equally reverse the logic and start from a manifestly
unitary amplitude, such as \ $\M_{\E}^{\mathrm{hard}}$ of Section~\ref{subsection:unitarity} or a Faddeev--Kulish amplitude,
which naturally breaks Lorentz invariance since a detector selects a preferred frame.
 Lorentz invariance, analyticity, and
crossing symmetry are then recovered in the detector scaling limit. We thank S.~Zhiboedov for discussions on this equivalent perspective.}

Stripped amplitudes can also be interpreted as representatives of an equivalence class of amplitudes that differ only by unresolvable soft radiation. 
Finite–apparatus effects are present but parametrically subleading, being
$O(\alpha, G, \E^2/M^2)$,
in contrast to our results which are valid at all-orders in $\alpha_{\E}$ or $G_{\E}$.
 \\

\noindent\textbf{Infrared–finite dispersion relations and bounds.}
Our framework yields dispersion relations that are fully IR finite and automatically circumvents problems related with the $1/t$ pole from photon or graviton exchange. The resulting constraints on the dispersive integrals---the ``arcs'' $\A_n$---persist in the presence of long–range forces and reduce smoothly to the standard short–range bounds when both $\alpha$ ($G$) and $\alpha_{\E}$ ($G_{\E}$) vanish. For instance, in QED the bounds take the form
\begin{equation}
    \A_n(g_*,\alpha_\E,\alpha)+O(\alpha)\ge 0\,,
\end{equation}
where $\A_n$ is a smooth function of the couplings, with calculable dependence on $\alpha_\E$ and on short–range interactions $g_*$. As illustration, in $\pi^+ \pi^+$ scattering with electromagnetism we have obtained explicit bounds on the leading EFT operator of the form, $
M^4\bar{c}_{2,0}+ (59.9\alpha_{\E})^2 +O(\alpha) \geq 0$, and analogous constraints for higher--dimensional Wilson coefficients, see \eqref{eq:boundsc31pi+pi+} and Fig.~\ref{fig:boundspipluspiplus}. Finite bounds for $\pi^+\pi^0$ scattering are reported in \eqref{eq:boundspi+pi0explicit} and \eqref{eq:fitpi0pi+}, see also Fig.~\ref{pi0pipmpivar} and \ref{pi0pipalphavar}. 

Similarly in gravity, $\A_n(g_*,G_\E,G)+O(G)\ge 0$, despite the familiar $1/t$ pole; for $\pi_0\pi_0$ scattering we have explicitly computed effects to order $O((GM^2 \log M/\E )^2)$, see \eqref{eq:boundsgr1}, \eqref{eq:boundsgr2} and Fig.~\ref{fig:boundsgr} and \ref{fig:boundsgr2}.

The key point is that we control the positivity properties of hard amplitudes that propagate no graviton (or photon) modes with $|\boldsymbol{q}|<\E$, and whose IR divergences have been stripped~off. 
As a result, potentially negative contributions from large--$\ell$ partial waves~\cite{Caron-Huot:2021rmr,Henriksson:2022oeu,Caron-Huot:2022ugt} 
remain finite, scale as $O(G)$ or $O(\alpha)$, see discussion in Section~\ref{sect:dispersion_relations}, and are parametrically subleading in the detector scaling limit.

The optimal detector, with optimal energy resolution $G_{\E}\to\infty$ or $\alpha_{\E}\to \infty$, is \emph{not}  optimal for constraining  irrelevant operators with positivity bounds. Operators less relevant than the minimal gravitational coupling are dominated by gravity in that limit: the leading gravitational logarithms %\eqref{eq:illustrativebound} 
can be resummed explicitly and the resulting bounds trivialize, effectively mirroring the reappearance of IR divergences. Equivalently, the gravitational time delay required to reach such an enormous apparatus swamps any microscopic causality--violating effect. Sensitivity to highly irrelevant Wilson coefficients instead requires a detector with finite $G_{\E}$. This is  possible because stripped amplitudes are IR finite. The optimal detector for bounding Wilson coefficients is a compromise between good energy resolution and acceptable time delay, see Fig.~\ref{quali}. 
This state of affairs could be improved further if we were able to control unitarity to subleading order in the scaling limit, which would allow access to even smaller detectors
that provide more stringent bounds while maintaining parametric accuracy.
This is something we hope to explore in future work. 

Remarkably, subleading tree-level $O(\alpha,G)$ contributions can also be controlled quantitatively, provided one assumes more about the UV completion. If the UV remains in a tree-level regime---namely, the high-energy amplitude is well approximated by a meromorphic function, as in large-$N$ theories \cite{Albert:2022oes,Fernandez:2022kzi,Berman:2023jys} or weakly-coupled stringy gravitational completions \cite{Camanho:2014apa}---then the residual error on the bounds is reduced to genuine loop effects, schematically $O(\alpha^2,G^2,\alpha g_*^2,G g_*^2,\alpha \alpha_{\E},\ldots)$, while the tree-level $O(\alpha,G)$ terms remain under control. This is stronger than simply assuming a weakly-coupled UV completion, which may still contribute through loops. It is really an assumption about the leading analytic structure of the amplitudes.  Moreover, such a statement only makes sense for IR-safe observables, such as the stripped amplitude $\M_{\E}$, since otherwise loop corrections would invalidate the tree-level approximation for the full amplitude $\M$ irrespective of the smallness of the couplings.\\

\noindent\textbf{Relation to previous literature.}
Our construction differs qualitatively from earlier attempts to incorporate long–range forces into positivity bounds. Previous analyses established nontrivial gravitational bounds only for $D>4$~\cite{Caron-Huot:2021rmr} or in AdS spacetime~\cite{Caron-Huot:2021enk}, with the bounds disappearing as $D\to4$ or the curvature is removed. In flat spacetime, working with IR--divergent amplitudes at fixed order in $G$ ($\alpha$) obstructs the existence of positive functionals: they must be integrable against the $1/t$ singularity, which fails in $D=4$. Introducing hard IR cutoffs by hand~\cite{Caron-Huot:2021rmr,Henriksson:2022oeu,Caron-Huot:2022ugt} does not resolve the issue: either the IR part of the dispersion relation diverges or the UV part becomes unconstrained. 
Moreover, the unitarity and analytic properties of the regulated amplitudes remain unclear
in this case, so as their meaning. Furthermore, as the IR cutoff is removed, the IR logarithms become large and
require resummation of all orders in $G$ ($\alpha$), as emphasized in \cite{Chang:2025cxc}, giving rise to apparently
paradoxical non--decoupling effects, as we discuss below.
Other approaches focusing solely on the graviton pole~\cite{Hamada:2018dde,Bellazzini:2019xts,Tokuda:2020mlf,Herrero-Valea:2020wxz,Noumi:2021uuv,Noumi:2022wwf} address only one aspect of the IR problem. 

In contrast, our framework removes the universal IR divergences from the outset by working with the IR--finite $\M_{\E}$, where the IR regulator $\epsilon$ has already been taken to zero.
The remaining scale, when exponentially smaller than all other hard scales, acquires a physical meaning in terms of detector resolution, see \eqref{eq:inclusiveXsection},  and should \emph{not} be taken to zero.
 In the detector scaling limit, potentially negative contributions remain finite and we are able to suppress them, thanks to  the small parameter $G/G_{\E}=1/\log(M/\E)\ll1$. This logic is fully analogous to the $1/N^2$ parametric suppression familiar from large--$N$
positivity bounds \cite{Albert:2022oes,Fernandez:2022kzi,Albert:2023jtd,Ma:2023vgc,Dong:2024omo},
in which loops, specific $u$--channel contributions, and  arcs at infinity in once--subtracted dispersions are systematically neglected. In our setting this is achieved by changing an experimental parameter $\E$ rather than $N$ and the theory itself. Moreover, while $G_{\E}/G\gg 1$, we can chose $\E$ such that   $G_{\E}$ is perturbative, avoiding the need to resum large logarithms which would effectively reintroduce IR divergences.

 Bounds derived in AdS space \cite{Caron-Huot:2021rmr} are conceptually closest to ours, in that the AdS curvature radius may be tentatively interpreted as an effective resolution scale $1/\E$,   rather than being taken to zero, although a precise recasting of AdS--bounds at finite radius in terms of flat space quantities has not been derived in the literature.  By contrast, extrapolation of  $D>4$ bounds to $D\to 4$ by holding $GM^2/(D-4)$ fixed, would provide technically finite results, but lack a physically meaningful and quantitative interpretation.

We also comment on the very interesting analysis of~\cite{Chang:2025cxc}, which
highlighted the necessity of resumming radiative effects in the vicinity of the
$t=0$ region for the full amplitude.
Indeed, fixed--order expansions in $G M^2/\epsilon$ are not meaningful once
$\epsilon\to0$, since the effective expansion parameter becomes large
independently of how small $G M^2$ is.
Ref.~\cite{Chang:2025cxc} further showed that, when the limits are taken in the
correct order, the $1/t$ pole is not by itself the fundamental obstruction to
positivity bounds in $D=4$, as one can construct positive functionals that
integrate the purely gravitational contribution to the dispersion relations to
finite values. For a similar perspective see also \cite{Blas:2020dyg,FuentesZamoro:2025exp}. 

At the same time, the analysis of \cite{Chang:2025cxc} led to apparently $G$--independent bounds---for
example, $g_2 M^4 + 259 \geq 0$ for amplitudes of the form
$\M = g_2 (s^2 + t^2 + u^2) + \cdots$.
This behavior suggested a puzzling lack of decoupling in the limit $G \to 0$,
since the unperturbed theory instead satisfies $g_2 \geq 0$.
We resolve this puzzle by observing that IR divergences suppress the full
amplitude at all kinematic scales, not only in the neighborhood of $t=0$ where
the $1/t$ contribution from graviton exchange arises.
As a consequence, bounds derived from IR--divergent amplitudes can formally be
written, but effectively reduce to trivial statements of the
type $g_2 M^4 \times 0 + 259 \geq 0$. 
This should be contrasted with our use of IR--finite stripped amplitudes, which
avoid this pathology altogether and yield finite bounds with a smooth
$G \to 0$ limit, see Section~\ref{sect:dispersion_relations} and explicit examples in Section~\ref{sec:positivitygravity}.  
Once IR divergences have been removed, and only
then, there is no need to work with the ``optimal resolution'' detector to efficiently constrain  irrelevant Wilson
coefficients.\\

In conclusion, long–range forces can be incorporated into the EFT–hedron without
disrupting the UV–IR link between low–energy EFTs and their microscopic completions. An especially
promising direction is the extension of the $S$–matrix bootstrap
\cite{Paulos:2016but,Paulos:2017fhb,Cordova:2019lot,EliasMiro:2019kyf,Homrich:2019cbt,Karateev:2019ymz,Guerrieri:2020bto,EliasMiro:2021nul,Guerrieri:2021ivu,Kruczenski:2022lot,EliasMiro:2022xaa,He:2023lyy,Guerrieri:2023qbg,Guerrieri:2024ckc,Eckner:2024ggx,Eckner:2024pqt,Gumus:2024lmj,He:2025gws},
which in $D=4$ currently operates only under the extreme assumption that both $\alpha$ ($G$) and
$\alpha_{\E}$ ($G_{\E}$) vanish, or the $S$–matrix becomes trivial. With stripped amplitudes, we can now move
away from this corner and allow for finite $\alpha_{\E}$ ($G_{\E}$), thereby incorporating the leading long–range
effects directly into numerical bootstrap analyses. This is particularly important in the case of electromagnetism for realistic applications of positivity and bootstrap constraints to particle phenomenology. Moreover, as already remarked, the future challenge is to develop a approach enabling to include corrections beyond the leading-log approximation, possibly based on IR--finite and physically relevant observables.

\subsubsection*{Acknowledgments}
We thank Jan Albert, Simon Caron-Huot, Tim Cohen, Miguel Correia, Stefano De Angelis, Ziyu Dong, Jaime Fernandez, Riccardo Gonzo, Kelian H\"aring, Carlo Heissenberg, Sebastian Mizera, Julio Parra--Martinez, Sara Ricossa, Javi Serra, and Sasha Zhiboedov  for useful discussions. J.B. is supported in part by Department of Energy grant DE-SC0007859 and by a Rackham Predoctoral Fellowship from the University of Michigan. G.I. is supported by the US Department of Energy under award number DE-SC0024224, the Sloan Foundation and the Mani L. Bhaumik Institute for Theoretical Physics.
F.R. is  supported by the SNSF under grant no. 200021-205016.  This research was supported in part by grant NSF PHY-2309135 to the Kavli Institute for Theoretical Physics (KITP).

\appendix

\section{Symmetric Dispersions}
\subsection{$tu$--symmetric dispersion relations}
\label{appendix:TUSymm}

We give here further details on the extension of $tu$--symmetric dispersion relations \cite{Bellazzini:2025shd} to the case of massive particles. These dispersion relations are used in the main text to bound $\pi^+\pi^+$ scattering in Section~\ref{sec:pipluspiplus}.

We consider stripped amplitudes satisfying
\begin{equation}
    \M_{\E}(s,t)=\M_{\E}(s,u=4m_{\pm}^2-s-t) \,.
\end{equation}
with $m_{\pm}$ the mass of the scattered particles.
It is then natural to adopt symmetric variables
\begin{equation}
\label{eq:tusymmvars}
s\,, \qquad q^2\equiv \frac{tu}{s} \,\,
\end{equation}
in terms of which we re-express the amplitude as
\begin{equation}
    {\M}_{\E}(s,q)\equiv \M_{\E}(s,t=t_{+}(s,q)) \,,
\end{equation}
where $t_{\pm}(s,q)$ are obtained by inverting Eq. \eqref{eq:tusymmvars}
\begin{equation}
\label{eq:inversetu}
t_{\pm}(s,q)=-\frac{s-4m_{\pm}^2}{2}\left(1\mp\sqrt{1-\frac{4q^2s}{(s-4m^2_{\pm})^2}}\right)\,, \qquad -s-t_{\pm}+4m_{\pm}^2=t_{\mp}\,.
\end{equation}
The Regge scaling is respected, indeed at $q$ fixed and large $|s|$ we have
\begin{equation}
    \lim_{|s|\to+\infty}\frac{t_{+}(s,q)}{s}=0\;\;,\;\;\;\lim_{|s|\to+\infty}\frac{t_{-}(s,q)}{s}=-1 \,.
\end{equation}
Thus it is convenient to identify $t=t_{+}(s,q)$ and $u=t_{-}(s,q)$, though this distinction is purely conventional, since the stripped amplitude is $tu$--symmetric. In particular, the branch-cut introduced in the definition of $t_{\pm}$ does not appear in the stripped amplitude, as trespassing it just amounts to exchanging $t_+$ and $t_-$.

The condition for the CoM--frame scattering angles associated to $t_{\pm}(s,q)$ to be real and physical, at $s\geq M_s^2\geq4m^2_{\pm}$, is
\begin{equation}
    0\leq q\leq \frac{M_s}{2}\left(1-\frac{4m^2}{M_s^2}\right)\equiv q_{max} \,.
\end{equation}
Armed with these definitions, we define the smeared arcs as
\begin{equation}
\label{eq:arcIRtusymmapp}
{\A}_n[\psi,{\M}_{\E,++}]=
\int_{\E}^{q_{\max}} dq\,\psi(q)\oint_{C}\frac{ds}{2\pi i}\frac{{\M}_{\E,++}(s,t_+(s,q))}{s^{3+n}} \,.
\end{equation}
where for the sake of the presentation we focus on the stripped amplitude ${\M}_{\E,++}$ for $\pi_+\pi_+\to\pi_+\pi_+$ scattering. We can deform the integration contour in the UV and use Regge scaling and crossing symmetry, obtaining the following dispersive representation:
\begin{equation}
\begin{split}
{\A}_n[\psi,{\M}_{\E,++}]=
% \int_{\E^2}^{p_{\max}} dp\,\psi(p)&\left[\int_{M^2}^{+\infty}\frac{ds}{2\pi i}\frac{\text{Disc}\,\M_{\E,++}(s,t_+(s,p))}{s^{3+n}}+\right.\\
% &\left.+\int_{-\infty}^{s^*}\frac{ds}{2\pi i}\frac{\text{Disc}\,\M_{\E,++}(s,t_+(s,p))}{s^{3+n}}\right]=\\
\int_{\E}^{q_{\max}} dq\,\psi(q)&\left[\int_{M_s^2}^{+\infty}\frac{ds}{2\pi i}\frac{\text{Disc}\,\M_{\E,++}(s,t_+(s,q))}{s^{3+n}}\right.\\
&\left.-\int_{-\infty}^{s_*}\frac{ds}{2\pi i}\frac{\text{Disc}\,\M_{\E,+-}(4m_{\pm}^2-s-t_+(s,q),t_+(s,q))}{s^{3+n}}\right]
\,.
\end{split}
\end{equation}
The second integral picks up the physical discontinuity of the crossed  process, as for $q^2\geq0$ and $s\leq0$ we have $4m_{\pm}^2-s-t_+(s,q)\geq 4m_{\pm}^2$ and $t_+(s,q)\leq 0$. Therefore, we fix $s_*=\frac{(4m_{\pm}^2-M_u^2)M_u^2}{M_u^2+q^2}$ and perform the change of variables
\begin{equation}
    u=4m_{\pm}^2-s-t_+(s,q)\Rightarrow s(u,q)=\frac{(4m^2_{\pm}-u)u}{q^2+u}\;,\;\;\frac{\partial s(u,q)}{\partial u}=\frac{4m_{\pm}^2q^2-u(2q^2+u)}{(q^2+u)^2} \,,
\end{equation}
obtaining
\begin{equation}
\begin{split}
{\A}_n[\psi,{\M}_{\E,++}]=
\int_{\E}^{q_{\max}} dq\,\psi(q)&\left[\int_{M_s^2}^{+\infty}\frac{ds}{2\pi i}\frac{\text{Disc}\,\M_{\E,++}(s,t_+(s,q))}{s^{3+n}}\right.\\
&\left.+(-1)^n\int^{+\infty}_{M_u^2}\frac{du}{2\pi i}\mathcal{K}_n(u,q)\frac{\text{Disc}\,\M_{\E,+-}(u,t_u(u,q))}{u^{3+n}}\right]
\,.
\end{split}
\end{equation}
with the definitions
\begin{equation}
\begin{split}
    \mathcal{K}_n(u,q)&\equiv\frac{u(2q^2+u)-4m_{\pm}^2q^2}{(u-4m^2_{\pm})^2}\left(\frac{q^2+u}{u-4m^2_{\pm}}\right)^{n+1} \,,\\
    t_u(u,q)&\equiv t_{+}(s(u,q),q)=\frac{(4m_{\pm}^2-u)q^2}{q^2+u}\,.
\end{split}
\end{equation}
The $t_u(u,q)$ defines a physical scattering angle $\theta_u$ within the integration regime, such that $-1\leq \cos\theta_{u}\leq 1$. Clearly the same holds for the angle $\theta_s$ associated to $t_{+}(s,q)$ in the first integral. In particular, we have
\begin{equation}
    \cos\theta_s=\sqrt{1-\frac{4q^2s}{(s-4m_{\pm}^2)^2}}\;\;,\;\;\;\cos\theta_u=\frac{u-q^2}{u+q^2} \,.
\end{equation}
Now, as for fixed--$t$ dispersion relations, we focus on the leading piece of the Arcs ${\A}^{(0)}_n[\psi,{\M}_{\E,++}]$ and leverage hermitian analyticity of the stripped amplitude and functional unitarity \eqref{eq:functionalopticaltheorem} to express the UV integrals in terms of hard amplitudes. Moreover, we can again extend the $q$ integration range down to $0$. The resulting UV representation is
\begin{equation}\label{eq:tuA0hard}
\begin{split}
{\A}^{(0)}_n[\psi,{\M}_{\E,++}]=\lim_{\substack{\alpha\to 0 \\ \E \to 0 \\ \alpha\log(M/\E)\to \alpha_{\E}}}
\int_{0}^{q_{\max}} dq\,\psi(q)&\left[\int_{M_s^2}^{+\infty}\frac{ds}{\pi }\frac{\overline{\text{Im}}\,\M^{\mathrm{hard}}_{\E,++}(s,t_+(s,q))}{s^{3+n}}\right.\\
&\left.+(-1)^n\int^{+\infty}_{M_u^2}\frac{du}{\pi }\mathcal{K}_n(u,q)\frac{\overline{\text{Im}}\,\M^{\mathrm{hard}}_{\E,+-}(u,t_u(u,q))}{u^{3+n}}\right]
\,.
\end{split}
\end{equation}
Expanding in partial waves the imaginary parts of the hard amplitude, we can express Eq. \eqref{eq:tuA0hard} in the form of Eq. \eqref{eq:UVrepresentation0}, with measures
\begin{equation}\label{eq:tuPelln}
    \begin{split}
        P_{\ell,n}^{\psi,s}& = \int_0^{q_{\text{max}}}dq \,\psi(q) \,P_{\ell}\left(\sqrt{1-\frac{4q^2s}{(s-4m_{\pm}^2)^2}}\right)\,,\\[0.2cm] P_{\ell,n}^{\psi,u}& = \int_0^{q_{\text{max}}}dq \,\psi(q) \,\mathcal{K}_n(s,q^2)P_{\ell}\left(\frac{s-q^2}{s+q^2}\right)\,.
    \end{split}
\end{equation}
Alternatively,  Eq. \eqref{eq:tuA0hard} can be written with the simplified notation
\begin{equation}\label{eq:tubracket}
    {\A}^{(0)}_n[\psi,{\M}_{\E,++}]=\!\!\!\!\lim_{\substack{\alpha\to 0 \\ \E \to 0 \\ \alpha\log(M/\E)\to \alpha_{\E}}}\!\!\!
    \int_{0}^{q_{\max}} \!\!\!dq\,\psi(q)\left[\Bigg\langle\! \frac{P_{\ell}\left(\sqrt{1-\frac{4q^2s}{(s-4m_{\pm}^2)^2}}\right)}{s^n} \!\Bigg\rangle_{\!\!\!++}\!\!\!+\Bigg\langle \frac{\mathcal{K}_n(u,q^2)P_{\ell}\left(\frac{u-q^2}{u+q^2}\right)}{u^n} \Bigg\rangle_{\!\!\!+-}\right] \,,
\end{equation}
with the averages, defined with a positive measure, given by
\begin{equation} 
\begin{split}
\big\langle(\ldots)\big\rangle_{++}&\equiv\sum_{\ell\in\mathrm{even}}(2\ell+1)\int_{M_s^2}^{+\infty}\frac{ds}{2\pi s^{3}}|a^{++}_{\E,\ell}|^2(s)(\ldots)\,,\\
\big\langle(\ldots)\big\rangle_{+-}&\equiv\sum_{\ell}(2\ell+1)\int_{M_u^2}^{+\infty}\frac{du}{2\pi u^{3}}|a^{+-}_{\E,\ell}|^2(u)(\ldots)\,.
\end{split}
\end{equation}

\subsubsection*{Tree-level bounds}

For illustration, we derive here some tree-level constraints on $\pi_+\pi_+$ scattering. In this simple case the smearing is not needed and we can work directly with the finite--$q$ Arcs, where $q$ is given in Eq. \eqref{eq:tusymmvars}
\begin{equation}
\label{eq:tusymmmastertree}
    {\A}_n(q)=\Bigg\langle \frac{P_{\ell}\left(\sqrt{1-\frac{4q^2s}{(s-4m_{\pm}^2)^2}}\right)}{s^n} \Bigg\rangle_{\!\!\!++}+\Bigg\langle \frac{\mathcal{K}_n(u,q)P_{\ell}\left(\frac{u-q^2}{u+q^2}\right)}{u^n} \Bigg\rangle_{\!\!\!+-} \,.
\end{equation}
The tree-level amplitude in the EFT reads
\begin{equation}
\M_{++--} 
{\Big|}^{\mathrm{EFT}}_{\mathrm{tree}} = -8\pi\alpha\frac{s^2-(6m_{\pm}^2+q^2)s+8m_{\pm}^4}{q^2s}+\bar{c}_{0,0}+\bar{c}_{1,0}s+\bar{c}_{2,0}s^2+\bar{c}_{2,1}q^2s+\bar{c}_{3,0} s^3 +\bar{c}_{3,1} q^2s^2+\ldots \,. 
\end{equation}
Notice that each Arc depends only on a finite number of Wilson coefficients and that the $\alpha$--terms, due to photons exchange, never enter the Arcs. For instance, for $n=0$ and $n=1$ we have
\begin{equation}
    \begin{split}
        \A_{0}(q)&=\bar{c}_{2,0}+q^2\bar{c}_{3,1}+q^4\bar{c}_{4,2} \,,\\
        \A_{1}(q)&=\bar{c}_{3,0}+q^2\bar{c}_{4,1}+q^4\bar{c}_{5,2}+q^6\bar{c}_{6,3} \,.
    \end{split}
\end{equation}
If we expand in powers of $q$ the corresponding UV representations, obtained from the master formula \eqref{eq:tusymmmastertree}, we get sum rules for the Wilson coefficients
\begin{equation}
\label{eq:treelevelpi+pi+Sumrules}
\begin{split}
    \bar c_{2,0}&=\big\langle 1 \big\rangle_{++}+\left\langle \frac{u^3}{(u-4m_{\pm}^2)^3} \right\rangle_{+-} \,,\\[0.2cm]
    \bar c_{3,1}&=-\left\langle J^2\frac{s}{(s-4m_{\pm}^2)^3} \right\rangle_{++}-\left\langle \frac{u^2(J^2-3)+4m_{\pm}^2u}{(u-4m_{\pm}^2)^3} \right\rangle_{+-} \,,\\[0.2cm]
    \bar c_{3,0}&=\left\langle \frac{1}{s} \right\rangle_{++}-\left\langle \frac{u^3}{(u-4m_{\pm}^2)^4} \right\rangle_{+-} \,,\\[0.2cm]
    \bar c_{4,2}&=\left\langle \frac{J^2(J^2-6)s^2}{4(s-4m_{\pm}^2)^4} \right\rangle_{++}+\left\langle \frac{[J^2(J^2-10)+8]u+16m_{\pm}^2(J^2-1)}{4(u-4m_{\pm}^2)^3} \right\rangle_{+-} \,,
\end{split}
\end{equation}
and an infinite set of null constraints, such as
\begin{equation}
\begin{split}
    0&=\left\langle J^2(J^2-6)(J^2-20) \frac{s^3}{(s-4m_{\pm}^2)^6}\right\rangle_{\!\!++}  +  \left\langle \frac{J^2(J^2-2)\left[(J^2-15)u+36m_{\pm}^2\right]}{u(u-4m_{\pm}^2)^3} \right\rangle_{\!\!+-} \,, \\[0.2cm]
0&=\left\langle J^2(J^2-6)(J^2-20)(J^2-42)\frac{s^3}{(s-4m_{\pm}^2)^8}\right\rangle_{\!\!++} \!\!\! -  \left\langle \frac{J^2(J^2-2)(J^2-6)\left[(J^2-28)u+64m_{\pm}^2\right]}{u^2(u-4m_{\pm}^2)^4} \right\rangle_{\!\!+-} \,,
\end{split}
\end{equation}
with $J^2=\ell(\ell+1)$.
Without even resorting to the null constraints, we can read from the first three sum rules the following tree-level bounds
\begin{equation}
    \bar c_{2,0}>0\;\;,\;\;\;-\frac{\bar c_{2,0}}{M_u^2-4m_{\pm}^2}<\bar c_{3,0}<\frac{\bar c_{2,0}}{M_s^2}\;\;,\;\;\;\bar c_{3,1}\leq3\frac{\bar c_{2,0}}{M_u^2} \,.
\end{equation}

\subsection{$stu$--symmetric dispersion relations}\label{app:stu}
In this section we provide a brief review of the $stu$--symmetric dispersion relations for massless scattering \cite{Li:2023qzs}, which are used in Section~\ref{sec:positivitygravity} to obtain bounds on gravitational scattering.
It is convenient to work in the 
variables 
\begin{equation}
s(z,q^2)=-\frac{3 q^2 z}{1+z+z^2}\,, \quad t(z,p)=s(z\xi,q^2)\,, \quad u(z,p)=s(z\xi^2,q^2)
\end{equation}
where  $\xi=e^{2i\pi/3}$ and the amplitude is understood as
\begin{equation}
\mathcal{M}_\E (z,q)\equiv \mathcal{M}_\E(s=s(z,q^2), t=t(z,q^2))\,.
\end{equation}Requiring physical scattering angles, 
$-1\leq \cos{\theta}\leq 1$, translates into the following allowed range for the variable $q$
 \footnote{Extension of the integration region beyond $q_{\mathrm{max}}$ can be used by considering additional conditions on the spectral density, see \cite{Pasiecznik:2025eqc}. }
\begin{equation}
0\leq q\leq  \frac{M}{\sqrt{3}}\equiv q_{\text{max}}
\end{equation}
We define the IR arc as \cite{Li:2023qzs,Beadle:2025cdx,Chang:2025cxc,Pasiecznik:2025eqc}
\begin{equation}\label{eq:stuIR}
\A_n[\psi,\mathcal{M}_\E]=\int_{\E}^{q_{\text{max}}}dq\, \psi(q)\left[\,\frac{(-1)^n}{3^{3(n+1)}}\oint_{C_{\text{IR}}}\frac{dz}{4\pi i }\frac{(z^3+1)(1-z^3)^{2n+1}}{q^{4n+4}z^{3n+4}}\mathcal{M}_\E(z,q)\right]
\end{equation}
where the contour is represented for instance on Fig. 6 of \cite{Beadle:2025cdx}. 

The UV representation can be conveniently expressed in terms of the Mandelstam variable $s$ leading to the dispersion relation
\begin{equation}
\A_n[\psi,\mathcal{M}_\E] =\int_{\E}^{q_{\text{max}}}dq\, \psi(q)\left[ \int_{M^2}^\infty \frac{ds}{4\pi i}s^{-3n-4}(3q^2+2s)(q^2+s)^n \,\text{Disc}_s\,\mathcal{M}_\E(s,q)\right]
\end{equation}

In analogy to the dispersion relations at fixed $t$ and $tu$-symmetric, we focus on the leading term $\A_n^{(0)}[\psi,\mathcal{M}_\E]$, extend the $q$ integral to $0$ and express the UV representation of the Arcs in terms of hard amplitudes obtaining
% Requiring weak gravity at the cut-off scale\footnote{This choice is more dictated by practical convenience, and could be eventually relaxed if a suitable computational scheme is available.}, that is $GM^2\ll1$, the smeared crossing symmetric Arcs  admit an IR perturbative expansion
%\begin{equation}
%\label{eq:expandingfunctionalsgravity}
%\mathcal{C}_n[\psi,\M_{\E}](g_*, G_\E,G)= \mathcal{C}^{(0)}_n[\psi,\M_{\E}](g_*, G_\E)+ G M^2\, \mathcal{C}^{(1)}_n[\psi,\M_{\E}](g_*, G_\E)+\ldots  \,,
%\end{equation}
%with $G_\E=GM^2\log M/\E$ and $g_*$ finite. The leading term for $GM^2\ll GM^2 \log M/\E$  is formally obtained as
\begin{equation}\label{eq:csdisp}
\A_n^{(0)}[\psi,\mathcal{M}_\E] = \!\!\!\!\lim_{\substack{G\to 0 \\ \E \to 0 \\ GM^2\log M/\E\to G_{\E}}} \!\!\! \int_{0}^{q_{\text{max}}}dq\, \psi(q)\left[ \int_{M^2}^\infty \frac{ds}{2\pi}s^{-3n-4}(3q^2+2s)(q^2+s)^n\,\overline{\text{Im}}\,\M^{\mathrm{hard}}_{\E}(s,q)\right]
\end{equation}
Expanding in partial waves the imaginary part we reproduce Eq. \eqref{eq:UVrepresentation0} with only one channel contribution
\begin{equation}\label{eq:stuPelln}
    P_{\ell,n}^{\psi,s}= \frac{1}{s^{2n+1}}\int_0^{q_\text{max}}dq\,\psi(q) \frac{3q^2+2s}{(q^2+s)^{-n}}P_{\ell}\left(\sqrt{\frac{s-3q^2}{s+q^2}}\right)\,.
\end{equation}
%\begin{equation}
 %   {\A}^{(0)}_n[\psi,{\M}_{\E}]=\!\!\!\!\lim_{\substack{G\to 0 \\ \E \to 0 \\ G\log\E\to G_{\E}}} \!\!\!
 %   \int_{0}^{q_{\max}} \!\!\!dq\,\psi(q)\,\Bigg\langle \frac{3q^2+2s}{(q^2+s)^{-n}}P_{\ell}\left(\sqrt{\frac{s-3q^2}{s+q^2}}\right) \!\Bigg\rangle_{\!\!\!CS} \,,
%\end{equation}
%with the crossing symmetric average given by
%\begin{equation} 
%\big\langle(\ldots)\big\rangle_{CS}\equiv\sum_{\ell\in\mathrm{even}}(2\ell+1)\int_{M^2}^{+\infty}\frac{ds}{2\pi s^{3n+4}}|a_{\E,\ell}|^2(s)(\ldots)\,.
%\end{equation}

\section{Numerics}\label{appendix:num} 
We briefly outline the approaches used for the numerical analysis throughout the paper. We used two conceptually different, though technically very similar, approaches, one for $\pi^+\pi^0$ scattering, see Section \ref{sec:pipluspizero}, and one for $\pi^+\pi^+$, Section \ref{sec:pipluspiplus}, as well as $\pi^{0}\pi^{0}$, Section \ref{sec:positivitygravity}.

\subsection{$\pi^+\pi^0$ Scattering}
One feature of $\pi^+\pi^0$ scattering that does not appear in the other examples discussed in this paper is that the low energy expansion of the relevant amplitude contains infinite Wilson coefficients, as in \eqref{eq:suansatztree}. 
To bound individual coefficients, we construct smearing functionals, $\psi(q)$, from a set of basis functions,
\begin{equation}
\psi(q) =\big(q_\text{max} - q \big)^2 \sum_{i=0}^{k_\text{max}} a_i q^{2i}\, ,
\end{equation}
where $q_\text{max}$ is the maximum value allowed for $q$ in order for the scattering angle to be physical. For $\pi^+ \pi^0$ scattering this is $q^2_\text{max} = (M^2-m_\pm^2)^2/M^2$. 

\begin{figure}[t]
\begin{centering}
    \includegraphics[width=0.6\textwidth]{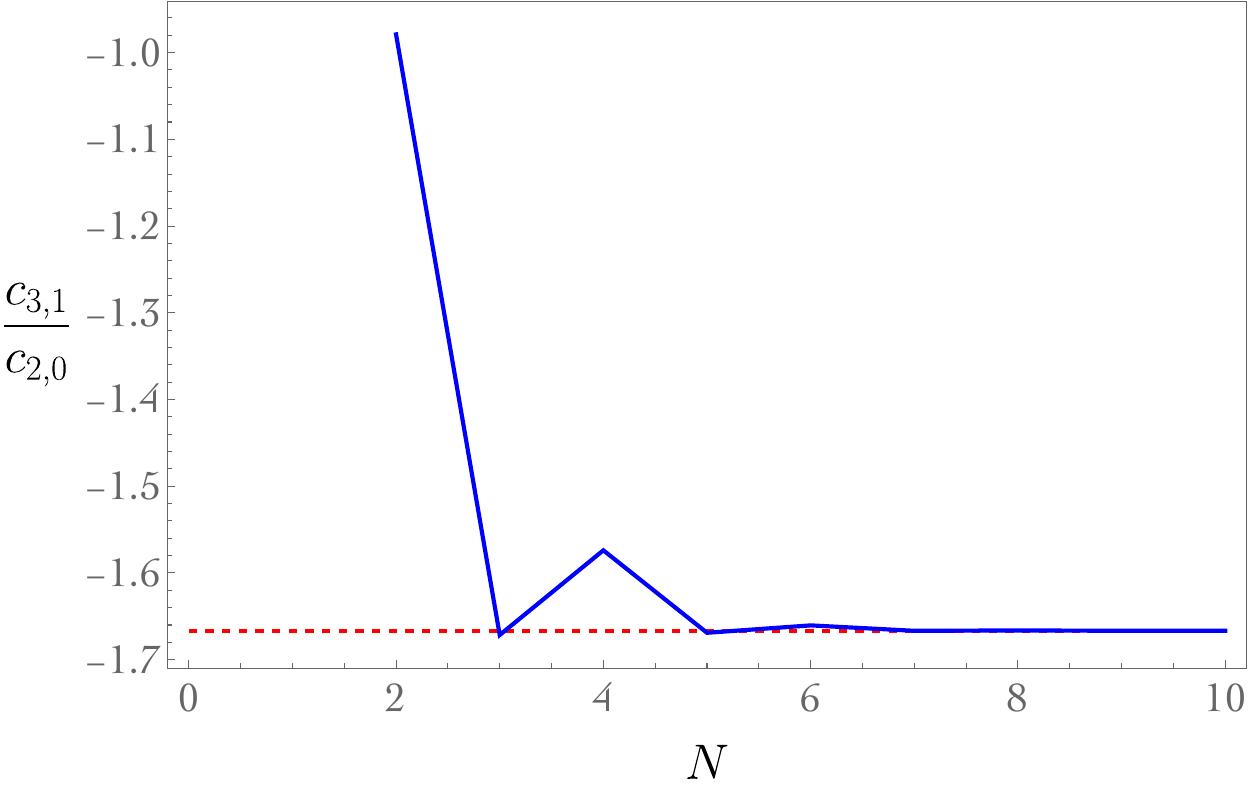}
     \caption{Convergence of the numerical bound as a function of the degree $N$ of truncation in the dispersion relations. The dashed red line corresponds to the analytic tree level bound given in \eqref{eq:boundspi+pi0tree} for $m^2_\pm=0.1$ in units of $M^2=1$.}
    \label{convoplot}
    \end{centering}
\end{figure}

Suppose we are working at tree level and we are interested in bounding only $c_{2,0}$ and $c_{3,1}$ in \eqref{eq:suansatztree}. Since the low energy expansion contains an infinite set of Wilson coefficients our logic is the following. Based on the ``EFT-informed'' assumption \eqref{eq:ass}, we truncate the full IR expansion at some degree $N$. We then choose $k_\text{max}$ (the parameter which controls how large a basis to use for the smearing functional) large enough such that we can enforce the conditions
\begin{equation}\label{eq:smearingconds}
\int_0^{q_\text{max}} \psi(q) q^{2i} \neq 0 \ \text{for} \ i=0,1  \quad,\quad \int_0^{q_\text{max}} \psi(q) q^{2i} = 0 \quad \forall \ i \in (2,N)\,.
\end{equation}
This removes any contribution from Wilson coefficients with powers of $q$ larger than two, i.e. any coeffients in the arc other than $c_{2,0}$ and $c_{3,1}$.
We then use semidefinite optimization \cite{Simmons-Duffin:2015qma} to optimize the $a_{i}$ coefficients to get bounds on the remaining Wilson coefficients ($c_{2,0}$ and $c_{3,1}$ in the example above). In order to have enough free parameters to remove contributions from the higher Wilson coefficients, get the optimal bound on the ones we do not kill, and require positivity in the UV, we observe that $k_\text{max} = N+2$. Finally, we evaluate the bounds as we increase $N$ (and $k_\text{max}$), and check if the bounds converge. For the tree-level amplitude in \eqref{eq:suansatztree}, the convergence of the bound $c_{3,1}/c_{2,0}$ as we increase $N$ is given in Fig.~\ref{convoplot}.

For the specific bounds in Fig.~\ref{pi0pipmpivar} and Fig.~\ref{pi0pipalphavar} we extend the logic above to include contributions from loops. Namely, we look for smearing functions such that 
\begin{equation}
\int_0^{q_\text{max}} L(q) \psi(q) q^{2i} \neq 0 \ \text{for} \ i=0,1  \quad,\quad \int_0^{q_\text{max}} L(q) \psi(q) q^{2i} = 0 \quad \forall \ i \in (2,N)\,
\end{equation}
where $L(q)=\M_{\E,+0}/\M^{\mathrm{tree}}_{\E,+0}$. 
% \mr{Are we repeating the same?}
% When we work with the expression which includes the effects of photon loops, the story is very much the same. The only difference is now, rather than using simply \eqref{eq:smearingconds}, we have to extract the full $q$ dependence multiplying whichever coefficient we wish to bound. Therefore, we use the conditions
% \begin{equation}
% \begin{split}
%     \int_0^{q_\text{max}} \psi(q) q^{2i}F(q^2,\alpha,\E,m_{\pm}^2,\mu^2) &\neq 0 \ \text{for} \ i=0,1\\ \int_0^{q_\text{max}} \psi(q) q^{2i}F(q^2,\alpha,\E,m_{\pm}^2,\mu^2) &= 0 \quad \forall \ i \in (2,N)\,\,,
% \end{split}
% \end{equation}
% where 
% \begin{equation}
% \begin{split}
%     F(q^2,\alpha,\E,m_{\pm}^2,\mu^2) = &\left[1+\frac{\alpha}{2\pi}\log\left(\frac{m_{\pm}^2(4m_{\pm}^2-t)}{\E^4}\right)\frac{t-2m_{\pm}^2}{\sqrt{t(4m_{\pm}^2-t)}}\arctan\left(\frac{\sqrt{t}}{\sqrt{4m_{\pm}^2-t}}\right)+\right.\\
%     &\left.+\frac{\alpha}{2\pi}\log\frac{4\pi\mu^2}{\E^2}\right]\,,
% \end{split}
% \end{equation}
% as in \eqref{eq:Mlambda1looppi0piplus}.
To make this numerically tractable, we choose small (but finite) values of $\alpha$ and $\E$, and have checked that the bounds are robust against making these parameters smaller as would be required by the strict scaling limit. In particular, we used $\alpha=10^{-3}$.

\subsection{$\pi^{+}\pi^+$ and $\pi^{0}\pi^0$ Scattering}
There are two main differences in the numerical evaluation of the bounds on $\pi^{+}\pi^0$ scattering and $\pi^{+}\pi^+$ scattering. First, in $\pi^{+}\pi^+$ scattering, the low energy arcs contain only finitely many Wilson coefficients. Therefore, we need not make the ``EFT-informed'' truncation and need only finitely many basis functions to completely isolate the contributions from the Wilson coefficients we are interested in bounding. For example, if we wish to bound $c_{2,0}$ at tree level, we can take the bounds on the zeroth arc, \eqref{eq:plusplusA0}, and use the freedom in the $a_{i}$ coefficients to simply set
\begin{align}\label{eq:pippipconds}
    \int_{0}^{q_{\max}}\psi(q)q^{2} = 0 \quad\text{ and } \quad \int_{0}^{q_{\max}}\psi(q)q^{4} = 0 \, ,
\end{align}
to get a valid lower bound on $c_{2,0}$. By increasing the number of included basis functions, we get strictly stronger bounds. In the reported bounds, we use four functionals.

The second qualitative difference between the $\pi^{+}\pi^+$ and $\pi^{+}\pi^0$ bounds is that the form of the zeroth arc \eqref{eq:A0approxlimit} is simple enough that we can analytically take the scaling limit $\alpha\to0$, $\E\to 0$, $\alpha\log M/\E \to \alpha_\E$. Choosing the following basis for the smearing functional,
\begin{align}
    \psi(q) = \sum_{k = 0}^{k_{\max}}\psi_{k}(q)\, , \quad\quad \psi_k(q) = \left(1-\frac{q}{q_{\max}}\right)^2~\frac{a_{n}q^{2k}}{q_{\max}^{2k}}\, 
\end{align}
we can analytically evaluate
\begin{align}
    \Psi_k^{n} = \lim_{\substack{\alpha\to 0 \\ \E \to 0 \\ \alpha\log(M/\E)\to \alpha_{\E}}} \int_\E^{q_{\max}}dq ~\psi_k(q) q^{n}
\end{align}
for $n = 2j$ for $j = 0,1,\ldots$ and 
\begin{align}
    \Psi_k^{\text{QED}} = \lim_{\substack{\alpha\to 0 \\ \E \to 0 \\ \alpha\log(M/\E)\to \alpha_{\E}}} \frac{8}{M^2(\log M/\E)^2}\int_\E^{q_{\max}}dq ~\psi_k(q) \frac{\log(q^2/4\pi\E)}{q^2}\,.
\end{align}
We can then write
\begin{align}
0 \leq \A_0[\psi,\M_\E] = \sum_{k=0}^{k_{\max}}a_{k}(\alpha_\E^2\Psi^{\text{QED}}_k+\bar{c}_{2,0}\Psi^0_k+\bar{c}_{3,1}\Psi^1_k+\bar{c}_{4,2}\Psi^2_k)\,.
\end{align}
Then, requiring \eqref{eq:pippipconds} ($\sum_{k}a_k\Psi^1_k = \sum_{k}a_k\Psi^2_k =0$) results in the bound on $\bar{c}_{2,0}$ \eqref{eq:boundc20pi+pi+}. Similarly, by making specific choices of $a_{0} \propto \psi'(0)$ and requiring only the second of the conditions in \eqref{eq:pippipconds}, we instead find the bounds \eqref{eq:boundsc31pi+pi+}. Fully numerical bounds showed in Fig. \ref{fig:boundspipluspiplus} are obtained by scanning values of $M^4\bar c_{2,0}/\alpha_{\E}^2$ and optimizing the bound on $M^4\bar c_{3,1}/\alpha_{\E}^2$.

The process is exactly the same for gravity, where we can again analytically evaluate the necessary integrals in the scaling limit. We use a slightly different set of basis functions which matches the choice of \cite{Caron-Huot:2021rmr,Chang:2025cxc,Beadle:2025cdx}.

\section{One-loop Amplitudes and Arcs}\label{app:amparcs}
We report in this Appendix the expressions for the one-loop amplitudes and the corresponding arcs.
\subsection{One-loop stripped amplitudes}\label{app:amps}

\subsubsection*{QED: $\pi_+\pi_+$}
\label{appendix:pi+pi+}

The one-loop stripped amplitude can be divided into two contributions, a purely electromagnetic one and another with insertions of the EFT Wilson coefficients. The  first is
\begin{equation}
    \begin{split}
        \M^{\text{1-loop}}_{\E,++}(s,t){\Big|}^{\mathrm{EM}}&=8\alpha^2        \left[\log\left(\frac{16\pi^2\E^4}{m_{\pm}^2(4m_{\pm}^2-s)}\right)+\right.\\
        &\qquad\qquad\left.+\frac{s-2m_{\pm}^2}{t}2\log\left(\frac{4\pi\E^2}{-t}\right)+\frac{s-2m_{\pm}^2}{4m_{\pm}^2-s-t}2\log\left(\frac{4\pi\E^2}{-4m_{\pm}^2+s+t}\right)\right]\times\\
        &\times\frac{s-2m_{\pm}^2}{\sqrt{s(-s+4m_{\pm}^2)}} \arctan\frac{\sqrt{s}}{\sqrt{-s+4m_{\pm}^2}}+\\
        &-8\alpha^2\left[\frac{s-2m_{\pm}^2}{t}\log\left(\frac{16\pi^2\E^4}{m_{\pm}^2(4m_{\pm}^2-t)}\right)+\frac{2m_{\pm}^2-t}{4m_{\pm}^2-s-t}2\log\left(\frac{4\pi\E^2}{-4m_{\pm}^2+s+t}\right)\right]\times\\
        &\times\frac{t-2m_{\pm}^2}{\sqrt{t(-t+4m_{\pm}^2)}} \arctan\frac{\sqrt{t}}{\sqrt{-t+4m_{\pm}^2}}+\\
        &-8\alpha^2\left[\frac{s-2m_{\pm}^2}{4m_{\pm}^2-s-t}\log\left(\frac{16\pi^2\E^4}{m_{\pm}^2(s+t)}\right)+\frac{-2m_{\pm}^2+s+t}{t}2\log\left(\frac{4\pi\E^2}{-t}\right)\right]\times\\
        &\times\frac{2m_{\pm}^2-s-t}{\sqrt{(4m_{\pm}^2-s-t)(s+t)}} \arctan\frac{\sqrt{4m_{\pm}^2-s-t}}{\sqrt{s+t}}\,.
        \end{split}\label{eqspleuf}
\end{equation}
Here we are keeping also the terms of order $\alpha_{\E}\alpha$, as they will be enhanced to $\alpha_{\E}^2$ by the smearing, because of the $1/t$ pole, while we are discarding genuinely $O(\alpha^2)$ terms. The second contribution is
\begin{equation}
    \begin{split}
         \M^{\text{1-loop}}_{\E,++}(s,t){\Big|}^{\mathrm{W}}\!\!\!\!&=\M^{\mathrm{tree}}_{++}\times\\
         &\times\frac{\alpha}{\pi}\left\{\left[\log\left(\frac{16\pi^2\E^4}{m_{\pm}^2(4m_{\pm}^2-s)}\right)\frac{s-2m_{\pm}^2}{\sqrt{s(-s+4m_{\pm}^2)}} \arctan\frac{\sqrt{s}}{\sqrt{-s+4m_{\pm}^2}}-(s\rightarrow t)-(s\rightarrow u)\right]\right.\\
        &\left.+\log\frac{4\pi\mu_{UV}^2}{\E^2}\right\}\,.
    \end{split}
\end{equation}
where again we neglected subleading contributions of order $O(\alpha^2)$.

\subsubsection*{Gravity}
The full $O(G^2)$ contribution to the one-loop gravitational amplitude is given by
\begin{equation}
    \begin{split}
        \M^{\text{1-loop GR}}_{\E}(s,t) &= -\frac{G^2\left(s^2+t^2+u^2\right)^2}{stu}\left[ s \log ^2\left(-\frac{s}{\hat \E ^2}\right)+t \log ^2\left(-\frac{t}{\hat \E ^2}\right)+u \log ^2\left(-\frac{u}{\hat\E ^2}\right)\right]\\
&-\frac{8 G^2 \left(s^4+t^4\right)}{s t} \left(\text{Li}_2\left(-\frac{u}{s}\right)+\text{Li}_2\left(-\frac{u}{t}\right)-\log \left(-\frac{t}{s}\right) \log \left(\frac{s}{t}\right)\right)\\
&-\frac{8 G^2 \left(t^4+u^4\right)}{t u} \left(\text{Li}_2\left(-\frac{s}{t}\right)+\text{Li}_2\left(-\frac{s}{u}\right)-\log \left(-\frac{u}{t}\right) \log \left(\frac{t}{u}\right)\right)\\
&-\frac{8 G^2 \left(s^4+u^4\right)}{s u} \left(\text{Li}_2\left(-\frac{t}{s}\right)+\text{Li}_2\left(-\frac{t}{u}\right)-\log \left(-\frac{u}{s}\right) \log \left(\frac{s}{u}\right)\right)\\
&-\frac{1}{15} G^2 \Bigg[\left(163\, s^2-283\, t u\right)\log \left(-\frac{s}{\hat\mu_{\text{UV}}^2}\right) +\left(163 \,t^2-283\, s u\right) \log \left(-\frac{t}{\hat\mu_{\text{UV}}^2}\right)\\
&+\left(163\, u^2-283\, s t\right) \log \left(-\frac{u}{\hat\mu_{\text{UV}}^2}\right)\Bigg]+\frac{1153}{25} G^2 \left(s^2+t^2+u^2\right)\,.
\end{split}\label{eqspleuf}
\end{equation}
Notice that the dependence in $\log^2{\E}$ cancels once momentum conservation $s+t+u=0$ is imposed. Therefore, the one-loop amplitude depends linearly in $\log{\E}$, as anticipated from the soft factor in Eq.~\eqref{eq:WeinbergEqualvanishingMasses}. The only terms which are relevant in the scaling limit \eqref{eq:generalscaling} are given by the first line.

The mixed term $O(G\hat c_{2,0})$ at one-loop reads
\begin{equation}
    \begin{split}
        \M^{\text{1-loop}}_{\E}(s,t){\Big|}^{\mathrm{GR}} &= -\frac{G\hat c_{2,0}}{2\pi}\Bigg[\frac{1}{150} \left(75 \pi ^2-6346\right) s t u-\frac{1}{900} \left(150 \pi ^2-4697\right) \left(s^3+t^3+u^3\right)\\
        &+\left(s^3-s t u\right)\log ^2\left(-\frac{s}{\hat\E^2}\right) +\left(t^3-s t u\right) \log ^2\left(-\frac{t}{\hat\E^2}\right)+ \left(u^3-s tu\right) \log ^2\left(-\frac{u}{\hat\E ^2}\right)\\
        &+\frac{1}{30} \log \left(-\frac{s}{\hat\mu_{\text{UV}}^2}\right) \left(181 s t u-76 s^3\right)+\frac{1}{30} \left(181 s t u-76 t^3\right) \log \left(-\frac{t}{\hat\mu_{\text{UV}}^2}\right)\\&+\frac{1}{30} \left(181 s t u-76 u^3\right) \log \left(-\frac{u}{\hat\mu_{\text{UV}}^2}\right)\Bigg]\,.\end{split}\label{eqspleufapp}
\end{equation}

\subsection{Various Arcs}

\subsubsection*{QED: $\pi_+\pi_+$}
We report here useful one-loop arcs, computed in the approximation $M\gg q_{max},m_{\pm}$. For the $n=0$ arc,  we get a contribution with one insertion of the Wilson coefficients
\begin{equation}
\begin{split}
    \delta\A_0^{(0)}[\psi,\M_{\E,++}]_{\text{Wilson}}&=\lim_{\substack{\alpha\to 0 \\ \E \to 0 \\ \alpha\log(M/\E)\to \alpha_{\E}}}\int_{\E}^{q_{max}} dq\,\psi(q)\mathcal{B}_0(q)  \,,
\end{split}
\end{equation}
with
\begin{equation}
\begin{split}
    \mathcal{B}_0(q)&=(c_{2,0} + 
 q^2 c_{3,1} + q^4 c_{4,2})\left[1+\frac{\alpha (2 m^2 + q^2)}{\pi\sqrt{-q^2 (4 m^2 + q^2)}}
  \arctan\left(\frac{\sqrt{-q^2}}{\sqrt{4 m^2 + q^2}}\right)  \log\left(
   \frac{16 \pi^2 \E^4}{m^2 (4 m^2 + q^2)}\right)\right]+\\
   &+ 2\alpha\frac{c_{1,0} + q^2 c_{2,1} - M^8 c_{5,0}/3 - 
    M^4 (c_{3,0} + q^2 c_{4,1} + q^4 c_{5,2})}{\pi M^2}\log\left(
   \frac{4 \pi \E^2}{M^2}\right)
    \,.
\end{split}
\end{equation}
The only--photons contributions is included in Eq. \eqref{eq:A0approxlimit}.

As for the $n=1$ arc the only--photon loop contribution is
\begin{equation}
\begin{split}
    \delta\A_1^{(0)}[\psi,\M_{\E,++}]_{\text{EM}}&=\lim_{\substack{\alpha\to 0 \\ \E \to 0 \\ \alpha\log(M/\E)\to \alpha_{\E}}}\int_{\E}^{q_{max}} dq\,\psi(q)
   8 \alpha ^2\frac{1-3 \log \left(\frac{4\pi\E ^2}{M^2}\right)}{3M^6}
    \,,
\end{split}
\end{equation}
which interestingly is of order $O(\alpha\alpha_{\E})$.
For the contribution with one insertion of the Wilson coefficients we get instead
\begin{equation}
\begin{split}
    \delta\A_1^{(0)}[\psi,\M_{\E,++}]_{\text{Wilson}}&=\lim_{\substack{\alpha\to 0 \\ \E \to 0 \\ \alpha\log(M/\E)\to \alpha_{\E}}}\int_{\E}^{q_{max}} dq\,\psi(q)\mathcal{B}_1(q)  \,,
\end{split}
\end{equation}
where
\begin{equation}
\begin{split}
    \mathcal{B}_1(q)&=(c_{3,0} + 
 q^2 c_{4,1} + q^4 c_{5,2})\left[1+\frac{\alpha (2 m^2 + q^2)}{\pi\sqrt{-q^2 (4 m^2 + q^2)}}
  \arctan\left(\frac{\sqrt{-q^2}}{\sqrt{4 m^2 + q^2}}\right)  \log\left(
   \frac{16 \pi^2 \E^4}{m^2 (4 m^2 + q^2)}\right)\right]+\\
   &+ 2\alpha\frac{c_{2,0} + q^2 c_{3,1}+q^4 c_{4,2} - 
    M^4 (c_{4,0} + q^2 c_{5,1})}{\pi M^2}\log\left(
   \frac{4 \pi \E^2}{M^2}\right)
    \,.
\end{split}
\end{equation}

\subsubsection*{Gravity}
The one-loop contribution to the $n=0$ arc with one Wilson coefficient insertion is obtained by integrating Eq. \eqref{eqspleufapp} against the $stu$-symmetric kernel \eqref{eq:stuIR}, and reads
\begin{equation}
\begin{split}
    \delta\A_0^{(0)}[\psi,\M_{\E}]_{\text{Wilson}}&=\lim_{\substack{G\to 0 \\ \E \to 0 \\ GM^2\log(M/\E)\to G_{\E}}}\int_{\E}^{q_{max}} dq\,\psi(q)\mathcal{C}_0(q)  \,,
\end{split}
\end{equation}
with
\begin{equation}
    \begin{split}
        \mathcal{C}_0(q)&= \frac{\hat c_{2,0}G}{2\pi}\Bigg[-\frac{ q^2 }{30M^2}\left(181+60 \log (\E^2/M^2)\right) \log \left(\frac{M^2}{ q^2}+1\right)\\
        &-4 \log (\E^2/M^2)-\frac{533 q^2}{20M^2}-\frac{136}{15}+2 \frac{ q^2 }{M^2}\,\text{Li}_2\left(-\frac{M^2}{ q^2}\right)-\frac{21q^2}{2M^2}  \log (\mu_{\text{UV}} /M^2)\Bigg]\,.
    \end{split}
\end{equation}

\addcontentsline{toc}{section}{References}

 \bibliography{bibs} 
 \bibliographystyle{utphys}

\end{document}